\documentclass[sigconf]{acmart}

\usepackage{float}
\usepackage{enumitem}
\usepackage{wrapfig}
\usepackage{dsfont}
\usepackage{xspace}
\usepackage[utf8]{inputenc}
\usepackage{wrapfig}
\usepackage{hyperref}

\usepackage{caption}
\usepackage{subcaption}

\PassOptionsToPackage{hyphens}{url}
\usepackage{amsmath, alltt, xspace, epsfig, graphicx, url, wrapfig, tabularx, color,time, stmaryrd, comment, diagbox, booktabs}
\usepackage{xcolor}
\usepackage{amsmath}

\usepackage{xcolor}
\usepackage{amsmath}
\definecolor{orange}{rgb}{1,0.5,0}

\newcommand{\mypara}[1]{\vspace*{0.1in}\noindent\textbf{#1 }}

\newcommand{\ignore}[1]{}

\ignore{

}

\newcounter{task}
\newcounter{thrust}

\usepackage{algorithm}
\usepackage{algpseudocode}

\newcommand{\bheading}[1]{{\vspace{4pt}\noindent{\textbf{#1}}}}

\newenvironment{packeditemize}{
\begin{list}{$\bullet$}{
\setlength{\labelwidth}{5pt}
\setlength{\itemsep}{0pt}
\setlength{\leftmargin}{\labelwidth}
\addtolength{\leftmargin}{\labelsep}
\setlength{\parindent}{0pt}
\setlength{\listparindent}{\parindent}
\setlength{\parsep}{0pt}
\setlength{\topsep}{0pt}}}{\end{list}}
\usepackage{etoolbox}
\usepackage{graphicx}
\usepackage{algorithm}
\usepackage{algpseudocode}
\usepackage[all]{nowidow}
\graphicspath{ {./figures/} }

\usepackage{wrapfig}
\usepackage{lipsum}

\usepackage{multirow}

\usepackage{siunitx}

\newcommand{\scheme}{\textsc{Styx}\xspace}
\newcommand{\pad}{\textsc{Pad}\xspace}
\newcommand{\pads}{\textsc{Pad}s\xspace}

\newcommand{\sys}{\textsc{Styx}\xspace}

\algnewcommand\algorithmicswitch{\textbf{switch}}
\algnewcommand\algorithmiccase{\textbf{case}}
\algnewcommand\algorithmicassert{\texttt{assert}}
\algnewcommand\Assert[1]{\State \algorithmicassert(#1)}%
\algdef{SE}[SWITCH]{Switch}{EndSwitch}[1]{\algorithmicswitch\ #1\ \algorithmicdo}{\algorithmicend\ \algorithmicswitch}%
\algdef{SE}[CASE]{Case}{EndCase}[1]{\algorithmiccase\ #1}{\algorithmicend\ \algorithmiccase}%
\algtext*{EndCase}%

\AtBeginDocument{%
  \providecommand\BibTeX{{%
    \normalfont B\kern-0.5em{\scshape i\kern-0.25em b}\kern-0.8em\TeX}}}

\renewcommand\footnotetextcopyrightpermission[1]{}
\acmConference[]{}{}{}

\title{\sys: Collaborative and Private Data Processing With TEE-Enforced Sticky Policy}

\begin{document}
\sloppypar

\author{Shixuan Zhao}
\affiliation{
\institution{The Ohio State University}
\city{Columbus}
\state{OH}
\country{USA}
}
\email{zhao.3289@buckeyemail.osu.edu}

\author{Weicheng Wang}
\affiliation{
\institution{Purdue University}
\city{West Lafayette}
\state{IN}
\country{USA}
}
\email{wang3623@purdue.edu}

\author{Ninghui Li}
\affiliation{
\institution{Purdue University}
\city{West Lafayette}
\state{IN}
\country{USA}
}
\email{ninghui@purdue.edu}

\author{Zhiqiang Lin}
\affiliation{
\institution{The Ohio State University}
\city{Columbus}
\state{OH}
\country{USA}
}
\email{zlin@cse.ohio-state.edu}

\begin{abstract}

Protecting sensitive information in data-driven collaborations, such as AI training, while meeting the diverse requirements of multiple mutually distrusted stakeholders, is both crucial and challenging. This paper presents \sys, a novel framework to address this challenge by integrating sticky policies with Trusted Execution Environments (TEEs). At a high level, \sys employs a hardware-TEE-protected middleware with a programming language runtime to form a sandboxed environment for both the data processing and policy enforcement. We carefully designed a data processing workflow and pipelines to enable a strong yet flexible data-specific policy enforcement throughout the entire data lifecycle and data derivation to achieve data-in-use protection, data lifecycle protection and dynamic collaboration. We implemented \sys and demonstrated its ability to make collaborative computing, such as joint AI training, more secure, privacy-preserving, and policy-compliant. Our evaluation shows the performance overheads imposed by \sys are reasonable on single-node computation with the capability to scale to a large distributed multi-node deployment.\looseness=-1
\end{abstract}



\everypar{\looseness=-1 }
\maketitle

\section{Introduction}
\label{chapter:SEDS}

We study the problem of protecting sensitive data used in collaboration among multiple stakeholders, where no single party is universally trusted. In this setting, while stakeholders are performing the data processing in a collaborative fashion, stakeholders are mutually distrusted and may have different yet specific policies on how their sensitive data can be used, what can be included in the output as well as how the output needs to be protected. This requires strong enforcement of data policies from every stakeholder throughout the entire processing as well as guarantees on those generated data. Three critical requirements emerge in such scenarios: First, \emph{data-in-use protection} must be achieved to ensure the computation itself does not violate the policies even if the computation is carried out by a party that is not universally trusted. Second, \emph{data lifecycle protection} is needed to ensure that any output must be allowed by the stakeholders' policies and is protected with proper output policies. Third, \emph{dynamic collaboration} must be supported to enable joint processing, ownership and policy decisions.\looseness=-1

\textit{Sticky policies} were introduced nearly two decades ago to support the denotation of the diverse policies of different stakeholders on different data by attaching data-specific policies to the data itself~\cite{karjoth2002platform,mont2003towards}. While sticky policies help forming a scheme to denote the data-specific policies, existing works focus on building access control with sticky policies~\cite{miorandi2020survey,Pearson2011sticky,pretschner2006distributed,padget2015policy}, meaning that enforcement schemes must be redesigned to achieve the three requirements we want to achieve.

At a glance, confidential computing seems to offer a promising approach to achieve data-in-use protection. It isolates sensitive data in a Trusted Execution Environment (TEE) during processing~\cite{cc-ibm,cc-arm,cc-intel,cc-techtarget} and has been widely adopted in cloud services, such as AWS~\cite{amazonIAM2011}, Google Cloud Platform~\cite{google}, and Microsoft Azure~\cite{cc-microsoft}. However, currently TEEs are primarily designed for outsourcing settings, where a single entity owns, processes, and stores the data. While TEEs offer remote attestation mechanisms to ensure the integrity of the applications even to multiple stakeholders, the attestation only conveys a hash and each stakeholder needs to verify the application before trusting the hash, which is particularly impractical in our setting where data can be processed by an untrusted third party using proprietary code. This means that TEEs fall short in providing data-in-use protection when the processing is done by other entities, not to mention data lifecycle protection in dynamic collaborations.\looseness=-1

To address these limitations, we propose \sys, a middleware framework that combines the data-specific policies offered by sticky policies with the strong enforcement of TEEs, and extends with runtime sandboxing to achieve secure collaborative sensitive data processing among multiple stakeholders. However, designing such a secure framework presents several challenges raised by the three critical requirements we discussed. One primary challenge is {ensuring data-in-use protection}, as traditional encryption and access control methods are inadequate once data is being actively processed. Another significant challenge is to {maintain data lifecycle protection}, which requires {policies to be enforced on the output side as well to specify not just what can be included in the output, but also what policies should be attached to the output just like the input data themselves}. Besides, achieving {dynamic collaboration} involves {consolidating different data policies from the input stakeholders and demands joint control and policies on the output}. In \sys, we employed a runtime sandboxing design to ensure that the I/O of the data processing program is under full control of the middleware framework, denying any unauthorized leakage and ensuring the data-in-use protection. We also designed a workflow with dual pipelines on the input as well as the output side to maintain the data lifecycle protection. The pipelines are made to take data from different stakeholders with respect to each data's policy on both the input and the output for the dynamic collaboration. We use the same runtime to allow pluggable policy engines to support fully customizable policies. 

We have implemented \sys into an architecture-independent framework and built a fully functional demo prototype upon it to demonstrate the feasibility and usability of this approach. The framework detached the protocol and framework from specific architectures, enabling data processing on heterogeneous processing nodes, which aligns with our vision for \sys in a distributed setting. The prototype of \sys is based on Intel SGX~\cite{vsgx} as the TEE and WebAssembly (WASM)~\cite{wasm} as the runtime. We were able to successfully port real-world applications, such as libonnx, an ONNX engine, into the system. We also evaluated \sys using a motivating example involving multiple hospitals collaboratively training a cancer classification model. Our evaluation results showed that \sys imposed comparable overheads on data access to conventional encrypted protocols such as HTTPS, while the runtime sandboxing design's extra overheads were also minimal compared to standard runtime performance.\looseness=-1

\mypara{Contributions.} 
We make the following contributions. 

\begin{packeditemize}
    \item \textbf{Sticky Policy and TEE Integration (\S\ref{sec:design}):} We present our novel approach \sys that combines the flexibility of sticky policy with the strong execution guarantee of TEE. We extended the integration to achieve data-in-use protection, data lifecycle protection and dynamic collaboration using our carefully designed input output pipelines, runtime sandboxing, pluggable policy engine support, addressing the challenge of enforcing sticky policies for both usage and data derivation in collaborative environments.
    \item \textbf{\sys Framework and Prototype (\S\ref{sec:implementation}):} We implemented \sys into an architecture-independent framework that contains the protocol and extendable interfaces. We built a fully functional prototype on Intel SGX with WASM, demonstrating the feasibility and practicality of \sys.
    \item \textbf{Application (\S\ref{sec:application}) and Evaluation (\S\ref{sec:performance}):} We applied \sys to our motivating example to showcase a real-world application usage on how \sys can protect the sensitive data from multiple stakeholders. We evaluated on both real-world applications and synthetic workloads as well as large-scale simulation and demonstrated a reasonable overhead.
\end{packeditemize} 

\section{Existing Approaches}
\label{sec:background}

\subsection{Cryptographic Approaches}

Various cryptographic approaches have been developed to offer data-in-use protection. Secure Multiparty Computation (SMC) techniques enable multiple stakeholders to jointly perform a computation while ensuring that each party only sees their designated output. The theoretical foundations for SMC were established in the 1980s and 1990s, with the garbled circuits approach for the two-party case~\cite{yao1982protocols,yao1986generate} and the secret-sharing based approach for the multi-party case~\cite{goldreich2019play,galil1987cryptographic}. These methods involve converting functions into boolean or arithmetic circuits then jointly evaluating them through a protocol. Recent research focused on improving the efficiency of these approaches~\cite{malkhi2004fairplay,huang2011faster,mohassel2017secureml,cramer2015secure,bogetoft2009secure,wang2023towards}. However, the performance overhead remains prohibitive for complex functions. Additionally, these methods offer limited protection for derived data. While they support different parties receiving different outputs, once an output is provided, other parties lose control over it even if their data was involved. Besides, requiring outputs to be encrypted using threshold cryptography significantly increases circuit size and overhead.

Homomorphic encryption \cite{gentry2009fully,brakerski2011efficient,vaikuntanathan2010computing,gentry2011implementing,dowlin2016cryptonets} allows computations performed directly on ciphertext without revealing plaintext. Only the party with the private key can decrypt the result. This is beneficial for single-entity scenarios where data storage and computation are outsourced to the cloud. However, it does not support multiple stakeholders each providing inputs and jointly controlling outputs. Furthermore, homomorphic encryption remains prohibitively expensive for complex computations.

\subsection{Access Control Approaches}

Another approach to protect data-in-use involves access control. Sandhu et al. \cite{xu2004applying} introduced the concept of usage control, which considers complex access requirements during data use and obligations after data usage. Languages for usage control were developed in a series of papers~\cite{park2002towards,park2004uconabc,zhang2005usage}, and enforcement of such policies was explored by augmenting SELinux MAC mechanisms~\cite{zhang2008security}. Byzantine-tolerant can also be employed~\cite{raptee,tnic}.

Researchers also explored protecting data during computation and transmission. The notion of \emph{sticky policies} was introduced almost two decades ago \cite{karjoth2002platform,mont2003towards} to help enterprises fulfill privacy policy commitments to end users. Sticky policies remain attached to data even as it moves across systems or networks. Recent surveys~\cite{miorandi2020survey,Pearson2011sticky} provide an overview of sticky policies. Similar ideas have also been introduced under \emph{distributed usage control}~\cite{pretschner2006distributed} and \emph{Policy-Carrying Data} (\emph{PCD})~\cite{padget2015policy}. The enforcement of sticky policy is difficult. Some researchers explored enforcing sticky policies using encryption techniques such as Identity-Based Encryption~\cite{boneh2001identity}, Attribute-Based Encryption~\cite{goyal2006attribute,bethencourt2007ciphertext}, and Proxy Re-encryption~\cite{green2007identity}. However, these techniques do not adequately support dynamic collaboration and derived data protection.

\subsection{TEE Approaches}

Trusted Execution Environments (TEEs) \cite{garfinkel2003terra,vasudevan2012trustworthy,murdoch2015introduction,tee-experimental} help protect data-in-use by creating secure, isolated environments that operate separately from the main operating system \cite{asokan2014mobile}. This isolation prevents unauthorized access or interference by other processes, ensuring data protection during the execution. Memory of TEEs is generally encrypted to maintain confidentiality, even if an attacker gains access to the system's physical memory \cite{kinney2006trusted}.\looseness=-1

Decisions about whether a program in the TEE can be trusted are done via remote attestations to verify the program's integrity and hardware genuineness, which offers no guarantees on the program's behavior, particularly on the input and output. Furthermore, TEEs face constraints due to limited input and output channels, posing challenges for collaborative computation. The software inside the TEEs is trusted universally after attestation, meaning that under a collaborative setting, one must fully trust the program to faithfully obey the policies blindly if the program cannot be formally verified or put under public inspection.\looseness=-1

Contemporary commercial cloud services incorporate confidential computing features using hardware-based security technologies, including enclaves and secure virtual machines. Examples include Amazon AWS Nitro Enclaves \cite{aws-nitro-enclaves} integrated with AWS Key Management Service~\cite{aws-kms}, and Google Cloud Platform's confidential shielded VMs \cite{Google2020ShieldedVMs} and Key Management Service \cite{Google2020KMS}. Hardware examples include Intel SGX~\cite{vsgx} and AMD SEV~\cite{sev2020strengthening}. These approaches support use cases where a single entity outsources computation to the cloud, instead of a collaborative computation design.

\section{Architecture Design} \label{sec:design}

\subsection{Motivating Scenario}
\label{sec:design:scenario}

\begin{figure}
\centering
\includegraphics[width=.8\columnwidth]{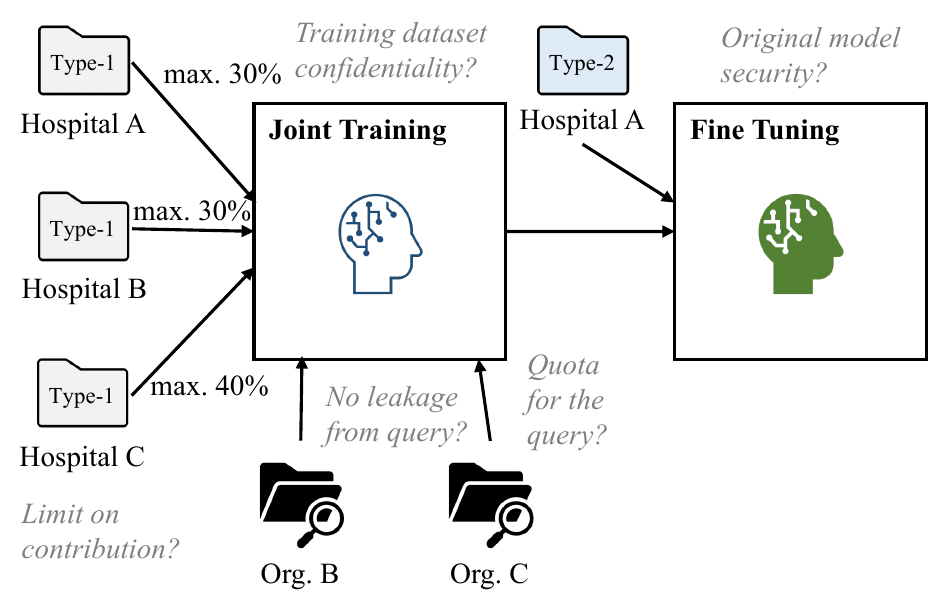}
\caption{A motivating scenario that hospitals jointly train a cancer classification model, fine-tune the model, and query the model.}
\label{fig:example}
\vspace*{-5pt}
\end{figure}

We first describe a scenario in which many organizations have sensitive data and would benefit from collaborating with each other to motivate our design requirements and design choices. For concreteness, we consider a medical use case in which multiple hospitals have collected diagnostic data from their patients and plan to jointly train a cancer classification model using machine learning (ML), as illustrated in~\autoref{fig:example}. To meet the regulation requirements and to protect patients' privacy, the diagnostic data must not be shared with each other. Note that information leakage can occur in several ways. First, information can potentially be leaked during training to the party controlling the software/hardware systems that run the training process. Second, information can be leaked through the model parameters, as information about the training data may be revealed through them. Third, information can be leaked through unlimited querying a model trained using the private diagnostic data and obtaining the result.  This last form of leakage may be unavoidable and therefore needs to be controlled rather than eliminated.\looseness=-1

We assume that hospitals have two types of diagnostic data, depending on their sensitivity and the regulations governing them. They are willing to contribute one type of data (type-1 data in \autoref{fig:example}) for the joint training of the ML model in a way that their data is not revealed, and the trained model parameters are not revealed directly to other hospitals. They also have another type of diagnostic data (type-2 data) that is more sensitive than type-1. They do not want type-2 diagnostic data to be used in training any model that other hospitals can query. However, they want to use type-2 diagnostic data to fine-tune the jointly-trained model to obtain a model that only themselves can use.  

In this setting, there is a jointly-trained base model. Some hospitals may not have any type-1 diagnostic data for training the base model, but are willing to pay money to either directly query the base model, or to fine-tune it using their type-2 diagnostic data. Some hospitals may have neither type-1 nor type-2 data and are willing to pay to query either the base model or some hospital's refined model. Furthermore, when a hospital queries an ML model, the hospital may require that the instances in the queries are not revealed to any other parties. \looseness=-1

This example illustrates the three requirements we described:

\bheading{Data-in-Use Protection.} We need to ensure that data used in computation are not revealed.  No one learns type-1 diagnostic data when training the base model.  No hospital learns the base model parameters while fine-tuning the model with its own type-2 diagnostic data.  A hospital submitting an instance to query a model does not learn anything other than the model prediction on the instance, and no other party learns anything about the instance.

\bheading{Data Lifecycle Protection.} We note that new data are being derived from computations involving protected data, and the derived data needs to be protected as well. For example, the base model is derived from type-1 diagnostic data, and fine-tuned models are derived from the base model and type-2 diagnostic data. The base model needs to be protected in a way that no hospitals can simply copy the model and use it in any way they want.  Similar controls need to be applied to the fine-tuned model.  The hospitals need to have confidence in these protections before the training can take place. However, it is challenging to support this.

\bheading{Dynamic Collaboration.}  The hospitals have policies determining under what conditions their data can be provided to train the ML model, and how the ML models can be used in the future. For example, hospitals may not be providing equal amounts of training data and want their shared rights to control the model to be determined by the amount of data.  Some hospitals may want to ensure that their share of data is sufficiently high so that they are able to exercise some higher level of control in the trained model. Some hospitals may want to ensure that their data do not constitute more than a certain percentage of the total training data. We need mechanisms to combine such constraints to determine whether the joint computation can be carried out.  Furthermore, not all usages for the derived data (ML models) may be known or can be determined at the time of training, and mechanisms are needed to update the policies later.  For example, initially the models may be only used by hospitals providing training data, and later it may be decided that the model can be sold for revenue with some revenue sharing scheme. \looseness=-1

\subsection{Design Considerations}
\label{sec:design:considerations}

We propose the following high-level design to support the key requirements of data-in-use protection, data-life-cycle protection, and dynamic collaboration. Specifically, our observation is that data's lifecycle follows a producer-consumer pattern. A program that generates protected data is called a data \textit{producer}.  A program that reads and uses protected data is called a data \textit{consumer}.  Some programs can be both a data producer and a data consumer. Contents of protected data, attributes about the data, and policies governing how the data can be used are encrypted and packaged together. We call such data \textit{Policy-Attached Data} (\pad). \looseness=-1 

However, ensuring the enforcement of the policy is not trivial. Sometimes a consumer's owner even has the motivation to peek at the data. There are three primary approaches to achieve this:

\mypara{Black-Box Approach.} A black-box approach requires the program to be verified and agreed by the stakeholders. The policy specifies which programs are allowed to access the protected data via hashes or certifications. The policy treats the program as a black-box and specifies nothing about the internal properties. This approach has the drawbacks of: (1) Has Limited support for dynamic collaboration since all stakeholders have to agree on the exact program; (2) Relies on program perfection for any protection to work; (3) Is impractical for complicated, diverse and proprietary consumer programs. \looseness=-1

\mypara{White-Box Approach.} A white-box approach uses the policies to determine what computation actions can be performed on protected data. The consumer program is no longer a general purpose program but works like a script to operate those allowed actions on the data. Such approach may be suitable for simple applications with limited routines, e.g., when training ML models using standard techniques. However, for general-purpose apps, the routines can be difficult to specify. Meanwhile, verification of these trusted actions can also be heavy when the actions are complicated or massive. Besides, this approach requires extra yet hard-coded design to ensure the output is properly protected.\looseness=-1

\mypara{Gray-Box Approach (Sandboxing).} This approach is a middle ground between the black-box approach and the white-box approach. The policy specifies constraints on the inputs and outputs of programs that can access protected data. To guarantee the enforcement of the policies, a special \textit{middleware} running inside a TEE is employed. The middleware sandboxes the programs that use protected data, and controls all the I/O of the program.  Before granting the program access, the middleware can verify that the policies of all provided input data are satisfied.  When the program generates outputs, the middleware ensures that correct policies are attached to the output against the input's policy. We choose to adopt this gray-box approach, as we believe it offers the best trade-off in terms of feasibility, flexibility, and trust in the data-access programs.\looseness=-1

\subsection{Threat and Adversary Model}
\label{subsec:threat-model}
\newcommand{\term}[1]{\ensuremath{\langle\mbox{#1}\rangle}}

\begin{figure*}
\centering
\small
\begin{tabular}{|rl|}
 \hline
\term{policy-attached-data} & ::= \term{metadata} \term{encrypted-payload} 
\\

\term{metadata} & ::= \term{data-id} \term{key-delegator-uri} 
\\



\term{plaintext-payload} & ::= \term{raw-data} \term{policy}+ \term{data-attribute}+
\\


\term{data-attribute} & ::= \term{attribute-name} \term{data-type-ref} \term{attribute-value} 
\\

\term{policy} & ::= \term{policy-lang-id} \term{input-constraints} \term{program-constraints} \term{output-constraints}  \\\hline

\end{tabular} 
\caption{A BNF specification for \pad. \term{encrypted-payload} is obtained from \text{aes-encrypt$_{\mbox{datakey}}$}(\term{plaintext-payload}).}\label{fig:bnf}
\vspace*{-5pt}
\end{figure*}

The ``gray-box'' approach allows us to define a very strong security model to protect the \pads. At high level, our threat model follows the same as common TEEs but with one significant difference: We consider data consumers to be untrusted. That is, other than those outside of the TEEs may want to steal the contents of \pads, the consumer program that processes the data may also want to leak the data or violate the policies attached to the data. We now discuss our threat model in details from external threats and internal threats.

\bheading{External Threats.} External threats refers to any adversaries outside the TEEs. They may get a \pad in the encrypted form but should not be allowed to read any portion of the data or the policy in the \pad in plaintext. We assume that TEEs are able to protect the code and data from external tampering to defend against these threats.\looseness=-1

\bheading{Internal Threats.} Contrary to conventional TEE threat models which uniformly trust everything inside the TEE, our threat model divides the software inside the TEE into two domains: the untrusted consumer program domain and the trusted infrastructure domain. The consumer program is considered as untrusted which can be buggy or even malicious. However, the infrastructure domain where the middleware, runtimes and libraries lie is trusted. The infrastructure code should provide a sandboxed environment to host the consumer program domain. While the consumer program is free to perform any computation inside the sandboxed domain, its I/O is completely controlled by the infrastructure. Whenever the consumer program would like to read in or write out any sensitive data, it must go through policy checks to ensure that such I/O is allowed. \looseness=-1

Our rationale is that unless being verified by a trusted third party or being open sourced for the community to inspect, a program cannot be considered as fully and universally trusted. Given that it is not practical to verify every version of every consumer program line by line and many consumer programs are proprietary, a consumer program naturally falls out of the trusted boundary.

\mypara{Out-of-Scope Threats.} Following common TEE threat models, this work does not address attacks against the TEE hardware stack (e.g., side-channel attacks~\cite{Standaert2010} and speculative-execution attacks~\cite{spectre}), or malicious consumers leaking data content via covert channels (e.g., leaking data by encoding the data with CPU load).

\subsection{Roles in Computation}
\label{sec:design:roles}

In addition to data producer and consumer as discussed, we have also introduced two roles called \textit{data custodian} and \textit{key delegator} to address the data ownership and cryptographic requirements. We now discuss the details of these roles:

\bheading{Data Custodian.} A data custodian owns certain nodes in the system (e.g., producers and/or consumers). It represents the data owner(s), instructs the producer on specific policies to be attached to the data and provisions encryption keys. One does not have to expect the data custodian to always be online.

\bheading{Key Delegator.} A key delegator is a service that stores encryption keys for data custodians and provisions these keys to validated producers and consumers. The only reason to have it is to help distribute the keys when the data custodian is not online and can be combined into the custodian as well. It is protected within a TEE and can be attested by the data custodian before being provisioned with the key. It will attest a target TEE before provisioning the encryption key to that TEE. Multiple instances can be launched and regular load balancing can be used for scalability. \looseness=-1

\bheading{Data Producer.} A data producer is a TEE-protected program that generates and packs the raw data into \pad. It is provisioned with policies and encryption keys directly from the data custodian.\looseness=-1

\bheading{Data Consumer.} A data consumer is a program that reads and operates on \pad. It is not allowed to access the data until the policy is checked and met. Even when it is granted with the access to the data under the policy, it must not leak any portion of the data directly. Its output should also comply with the original policy's restriction on the computation outputs.

\subsection{The Elements of Policy-Attached Data}
\label{sec:design:pad}

A BNF specification of policy-attached data (\pad) in \autoref{fig:bnf} to define how a \pad should be packed. To achieve the strong enforcement of the policy for a \pad, the \pad can only be accessed inside a TEE by a consumer after the policy is met. This requires most of the payload to be encrypted, leaving only minimal information in plaintext to help the TEE retrieve the key to decrypt the actual payload. \looseness=-1

\mypara{Metadata.} The \term{metadata} is the only plaintext portion, providing information for retrieving the decryption key for a \pad. \looseness=-1

Each \pad has exactly one data custodian who possesses the key used to encrypt the payload in \pad. The key is provisioned to the key delegator described in \autoref{sec:design:roles} whose URI \term{key-delegator-uri} is included in the \term{metadata}. Depending on the custodian's choice, the key can be data specific, and therefore a \term{data-id} is also included. When a consumer attempts to access the \pad, the trusted middleware can use these information to fetch the key from the key delegator after a successful attestation. \looseness=-1

\mypara{Payload.} The \term{plaintext-payload} is a sequence of three elements: the \textit{raw data}, one or more \textit{policies} and the \textit{data's attributes}.  Intuitively, each policy specifies that the \pad can be read by a consumer program that complies with the policy written in the policy language of the given \term{policy-lang-id}, and used as the input identified by \term{input-constraints}, provided that the program satisfies the \term{program-constraints}. The output of the consumer can be restricted with the \term{output-constraints}, which can specify what is allowed in the output as well as what policy or ownership should be attached to the output. The \term{data-attribute} element is used to store information specific about the data that is used during the policy evaluation (e.g., time of creation, data count). \looseness=-1

The payload is encrypted with an optional Cryptographic Message Authentic Code (CMAC) for integrity check before being appended to the metadata to form a \pad.

\mypara{Policy Language.} Each policy is written in a specific \textit{policy language}, referred with the \term{policy-lang-id}. The policy is to be interpreted by a policy engine implemented as a trusted pluggable module containing no data and can be made open for inspection. The engine makes Boolean decisions based on the input, the consumer program and the output against the policy. We discuss more about the policy engine in \autoref{sec:design:arch}.

\subsection{Workflow and Components of \scheme}
\label{sec:design:arch}

\begin{figure*}[ht]
    \centering
    \includegraphics[width=.75\textwidth]{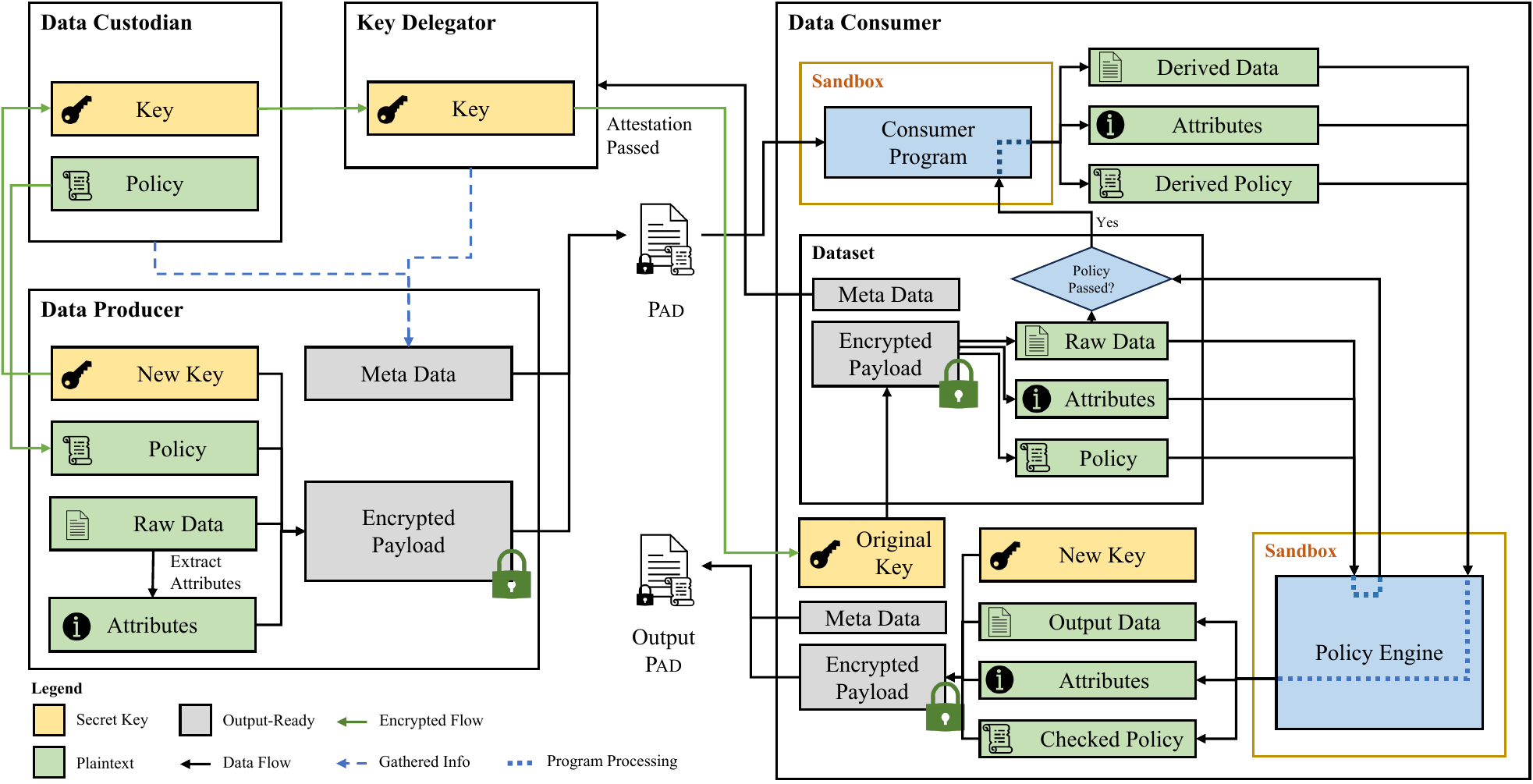}
    \caption{The workflow of the \scheme.}
    \vspace*{-5pt}
    \label{fig:workflow}
\end{figure*}

\begin{figure}[t]
    \centering
    \includegraphics[width=\linewidth]{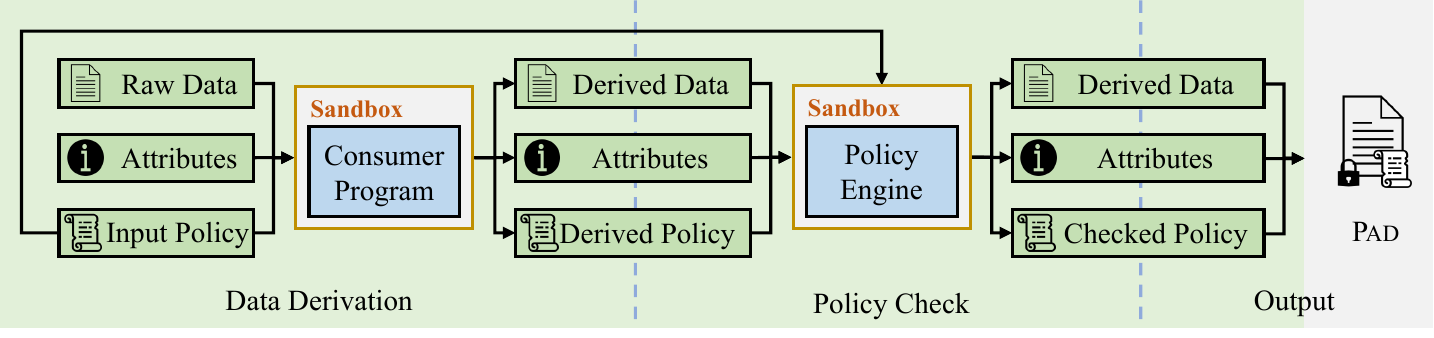}
    \caption{Data and policy derivation workflow. Green background means trusted domain. Gray background means untrusted domain (sandboxes or outside of TEE).}
    \label{fig:deriv-workflow}
    \vspace*{-10pt}
\end{figure}

The workflow and architecture of \scheme must enforce the policy even if a consumer program is not trusted and can be compromised. We designed a complete workflow and interaction between the roles and components and illustrated them in \autoref{fig:workflow}.\looseness=-1

\bheading{Preparation.} First, the data custodian needs to specify the key delegator, the policies attached to the data, and data attributes used in policy verification before the data is created to the data producer. For some cases, the producer's infrastructure is also owned by the data custodian, and such an agreement can be hard-coded in the producer, e.g., an IoT device serves as a producer and the device owner is also the data custodian. If the producer's infrastructure is not directly owned by the data custodian, a secure communication is required before data is created. \looseness=-1

\bheading{Data Creation in the Producer.} The producer first generates the raw data or obtains the raw data from trusted channels of the data custodian. After extracting data attributes needed by the policy, it then builds and encrypts the payload with a randomly generated symmetric data encryption key inside the TEE. This data encryption key is sent back to the data custodian who will provision it to the key delegator. The generated \pad can now be sent to the outside.

\bheading{Data Computation in the Consumer.} The critical protection needed in our design happens in the consumer that uses the data and derives new data based on the input. To solve the challenge that the consumer program is not considered as trusted, we propose to use runtime sandboxing technique to achieve a full control over the consumer program's I/O, forcing any I/O to go through our designed pipelines that enforce policy checks. This means that the TEE environment will run a consumer middleware, in which the consumer program is executed inside a language runtime. This middleware contains no proprietary logic and can be open for inspection for trust. There are four major components of the middleware:\looseness=-1
\begin{packeditemize}
    \item \textbf{Runtime.} The runtime serves as the sandboxed environment for the consumer. Technically, any runtime that requires I/O to be handled by the runtime can be used (e.g., Python, WebAssembly). The consumer program is then written or compiled into the language or bytecode of the runtime for execution.
    \item \textbf{\scheme Framework.} We designed our I/O pipelines and other supporting infrastructures into an architectural-independent framework library. The framework offers APIs to adapt to different runtimes, cryptographic algorithms as well as TEE interfaces.
    \item \textbf{Dataset.} Input \pads must form a \textit{dataset} before a policy check. This is because policies can have inter-data dependencies such as percentage limit of one entity's data in the entire dataset.
    \item \textbf{Policy Engine Support.} As described in \autoref{sec:design:pad}, the policy engine is a trusted pluggable module that can support sophisticated policy languages to restrict the input, program and output. Essentially, this policy engine should be able to perform complex logics but eventually needs to make a \textit{Boolean} (``\textit{yes or no}'') \textit{decision} before the raw data from the \pads being made accessible to the consumer as well as before derived data packed in \pads being allowed for output. To achieve this, we uses the same runtime infrastructure to allow policy engines be written in the same fashion as the consumer program. For a policy engine, it has to implement 2 policy check calls for the input and output pipelines. The input call will be called by the middleware before the consumer program tries to access the data. It takes the \pads in the dataset and the program as the input and returns a Boolean value for if the consumer program is allowed to access the data based on the \term{input-constraints} and \term{program-constraints} in the policies of these \pads. The output call will be called by the middleware when the consumer program requests an output. It takes the derived raw data, attributes and policy as well as the policies from the \pads in the dataset then returns a Boolean value for if the consumer program is allowed to output the derived data based on the \term{output-constraints}. The corresponding pipeline will then allow or deny the input/output based on the returned Boolean status from the policy check call of the policy engine. An example policy engine design will be presented in \autoref{sec:application}.
\end{packeditemize}

With this middleware, the burden of trust is now onto the middleware instead the consumer program. When the consumer program would like to read certain data, it must first gather all the \pads it needs to access and form the dataset. The dataset infrastructure will perform the decryption by contacting the key delegators of the \pads. Key delegators will perform a remote attestation to the middleware using hardware TEE's remote attestation mechanism to ensure that the middleware is genuine, and then provision the keys to the dataset infrastructure. After decrypting the \pads, the dataset infrastructure will run a policy check across all \pads using the policy engine. It will only provide raw data access to the consumer program if the policy check is passed on every \pad.

For data derivation, since policies may have output constraints, whenever the consumer program attempts to output any derived data, it must go through the output pipeline as illustrated in \autoref{fig:deriv-workflow}. Any other output interfaces are disabled. The consumer program needs to act like a data producer to generate proposed data attributes and derived policy including custodian info for joint ownership. The data, attributes and policy will go through the policy engine to for compliance check against the input \pads' policies. Only if the output data, attributes and policy passes the check before they can be packed into a new \pad. Just like a producer, the new key will be sent to the new \pad's custodian and then the key delegators for further distribution.

\section{Implementation}\label{sec:implementation}

In this section, we provide details of our prototype implementation of \scheme. We implemented a fully-functional system based on Intel SGX~\cite{vsgx} and WebAssembly~\cite{wasm} to demonstrate the feasibility and usability of our design. We have also successfully ported real-world applications into the system, including libonnx, an ONNX engine. The implementation consists of 13470 lines of code. Note that while we chose WebAssembly and SGX for this demo, the runtimes and TEEs are not limited to them and and can be swapped with other alternatives based on the need of the application. The source code of \sys will be released on GitHub upon paper publication.

We follow our design philosophy that the data protocol and framework should be detached from a specific architecture so the data can be processed on heterogeneous processing nodes. Then on top of this, we build the demo system.

\subsection{\pad Format}

\begin{figure}[t]
    \centering
    \includegraphics[width=.35\textwidth]{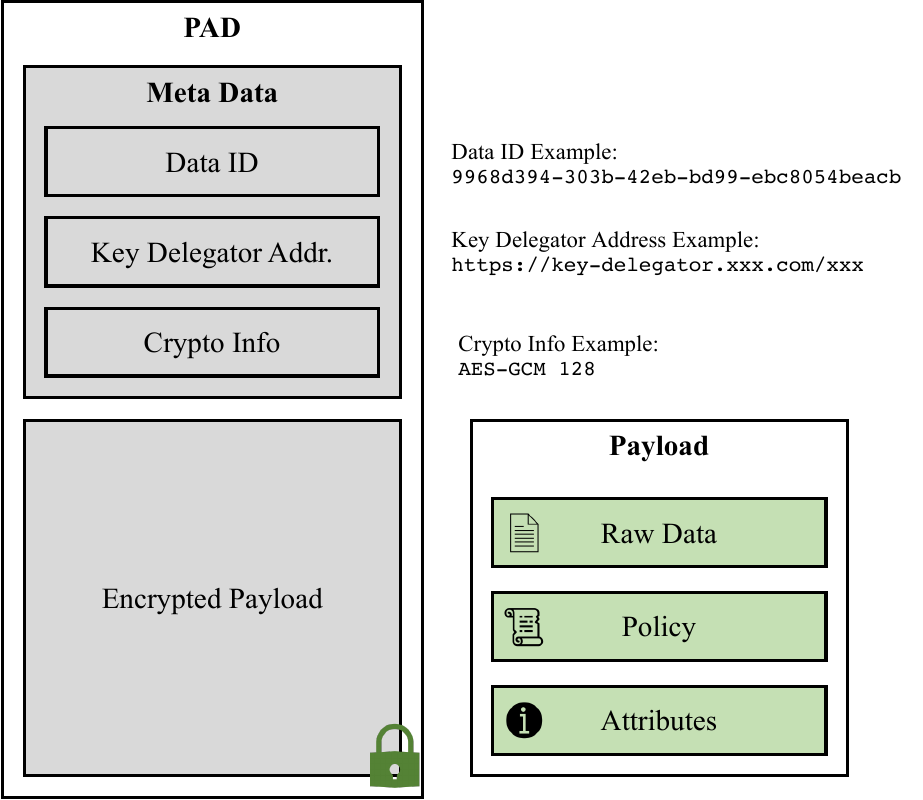}
    \caption{The format of \pad in our implementation. The left side shows the view of a \pad before decryption.}
    \label{fig:pad-struct}
    \vspace*{-10pt}
\end{figure}

The \pad format essentially forms the data protocol that must be both architecture independent and secure. We followed the BNF presented in \autoref{fig:bnf} for the layout but added and specified a few fields in details. For all IDs used in \pads, we chose UUID to ensure that the ID is universally unique. We also added a ``Crypto Info'' field to represent the cryptographic algorithm used. This is to enable heterogeneous computation, particularly for smaller IoT devices with limited choices. To ensure both confidentiality and integrity, a cryptographic algorithm with CMAC can be used (e.g., AES-GCM).

\subsection{Framework}

\begin{figure}[t]
    \centering
    \includegraphics[width=.34\textwidth]{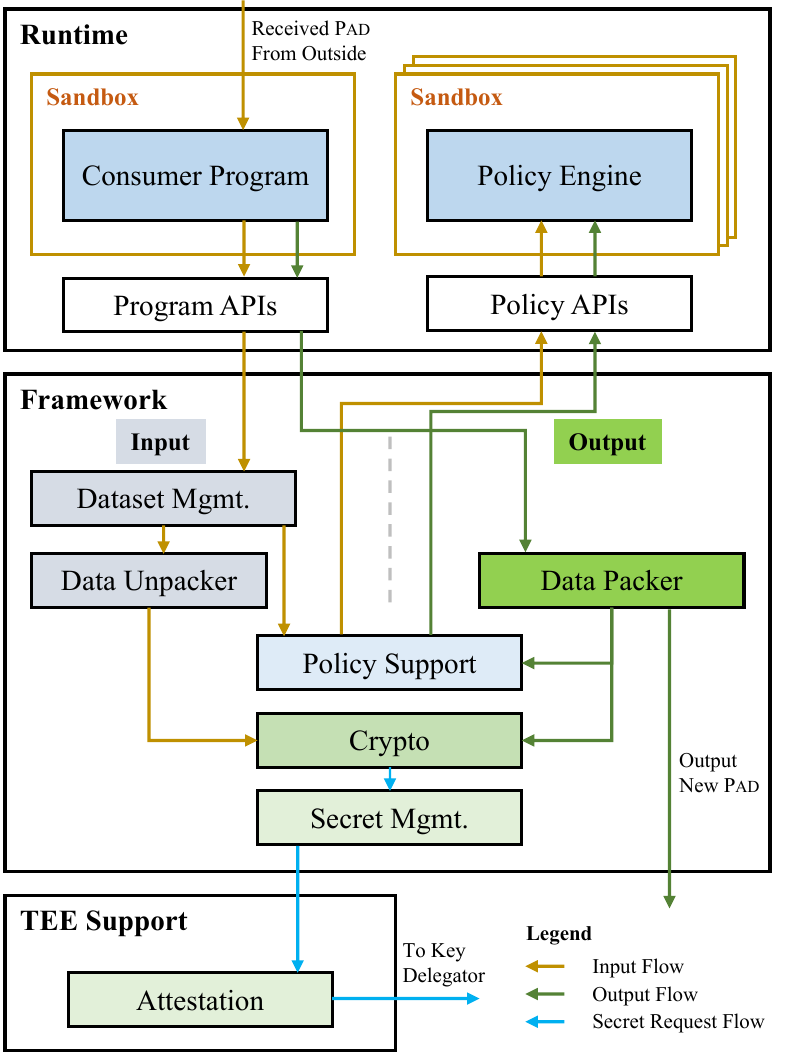}
    \caption{A consumer middleware hosting multiple policy engines and a consumer program.}
    \label{fig:middleware-impl}
    \vspace*{-10pt}
\end{figure}

We implemented our protocol into a framework library with configurable modules based on the role of the target program so that every role can use the same framework code base. There are 6 modules in the framework: \textit{Data Packer}, \textit{Dataset Management}, \textit{Data Unpacker}, \textit{Cryptographic Interface}, \textit{Secret Management}, \textit{Policy Engine Support}, \textit{TEE Support}, \textit{Runtime Support}. While different roles require different modules, a consumer middleware requires all of them and can better demonstrate the architecture on a single node. We therefore illustrate a consumer middleware architecture in \autoref{fig:middleware-impl} and discuss each component in detail below. \looseness=-1

\bheading{Data Packer.} The data packer provides interfaces for a program to pack data and policy into a \pad. It is needed by both producers and consumers. When the program wants to pack a piece of data with its policy, it provides the data packer with the elements inside the payload and the metadata. If the data packer is working in a consumer middleware, it will pass the raw data and proposed policy to the policy engine before packing the data. The packer will inquire the secret management to retrieve or generate the key, encrypt the payload and produce the \pad.  \looseness=-1

\bheading{Dataset Management.} The dataset management module is needed by the middleware of the consumer to build a one-shot dataset over which the policy is evaluated before granting the access to the consumer program. When the program wants to access a set of \pads, it will first have to create a dataset, add \pads into it and then request a policy check over the entire dataset before it can be allowed to access the data in the dataset.  \looseness=-1

\bheading{Data Unpacker.} The data unpacker is used by the dataset management module to extract data and policy from a \pad. When a \pad is added into the dataset, the data unpacker will inquire the secret management module to either retrieve the key locally or to fetch the secret key from the remote key delegator and then decrypt the payload. The plain-text payload is sent back to the dataset management module. Note that the program cannot access the plain-text payload until the policy check is passed.\looseness=-1

\bheading{Cryptographic Interface.} To adapt to the diverse yet possibly limited cryptographic capabilities of different platforms, we designed a unified interface for cryptographic so that developers can easily add or configure the algorithms they want to support. This module is required on both producer and consumer. While optional, integrity check requires the algorithm to generate a CMAC.

\bheading{Secret Management.} The secret management module is one of the fundamental security gatekeepers of our design. It is required on all roles and serves as a mapping from the data ID to the data's key (the secret). The secret can be loaded locally or fetched from the remote key delegator. The custodian of the data has the ability to push the data encryption key to the remote key delegator to make the key available to other programs. When the remote key delegator receives a fetch/push request, it will perform mutual attestations against the client connecting to it to prove its identity as well as ensuring the client is using an approved infrastructure (e.g., a genuine consumer middleware) on a trusted hardware.

\bheading{Policy Engine Support.} Each policy language has a policy engine. The engine is compiled into a software module that the runtime can execute. A policy engine module is loaded and identified by the policy type UUID. The policy evaluation interfaces are exposed to the dataset management and the data packer to check for compliance. We support multiple policy languages for different data by loading the corresponding policy engines.\looseness=-1

\bheading{TEE Support.} In our design, all roles are protected within a TEE. Our framework can support different TEEs as long as the corresponding attestation and communication interfaces are implemented. The attestation must be mutual, meaning that when the attestation is completed, both sides can trust each other. The attestation should also exchange a ephemeral communication key so that the traffic between the two TEEs are encrypted. This channel is only necessary for secret management. \looseness=-1

\bheading{Runtime Support.} The consumer program is sandboxed within a programming language runtime (e.g., WebAssembly, Python, etc.) to prevent it from leaking data unintentionally or deliberately. \sys also uses the same runtime to support policy engines. In our implementation, we specified a set of interfaces to manage the runtime and programs' lifecycle. To add support of a new runtime, a developer only needs to bind the runtime's interfaces to our interfaces for the framework to use the new runtime.  \looseness=-1

\subsection{Integration with Intel SGX and WAMR}
\label{subsec:impl-complete}

We built a demo prototype of \sys with Intel SGX as the TEE~\cite{shinde2017panoply} and the WebAssembly Micro Runtime (WAMR)~\cite{wasm} as the runtime.\looseness=-1

\bheading{Integration with Intel SGX.} Intel SGX provides an isolated and protected enclave environment within the user space memory. The enclave can access the data outside but the outside cannot access the data inside. This means that we can handle plain-text data inside the enclave without leaks. Intel SGX supports hardware-based remote attestation that verifies the integrity of the loaded software and the genuineness of the hardware. The attestation flow also creates an encrypted communication channel between the two enclaves using DH key exchange algorithm for the secret management module.\looseness=-1

\bheading{WAMR and Programs in the Middleware.} WebAssembly (WASM) is a low-level IR that can be interpreted or even compiled natively into machine code. It can be sandboxed to its own memory space and its I/O is controlled by the runtime. Since WASM can be generated from LLVM IR, any programming language that LLVM supports can be compiled into WASM, allowing great flexibility for developers. WAMR is an embedded runtime that has full WASM support while being natively integrable into SGX.\looseness=-1

To build a program using \sys, a developer can simply write a program that LLVM supports (e.g., C++) and call the interfaces we provided that are in pure C. In our design, for a program, we only export the dataset management interfaces to it and all the \pad accesses from the program must go through the dataset lifecycle by loading, checking policy and finally accessing. The I/O interfaces of the WAMR can be limited or even removed to prevent the program from leaking data without protection.\looseness=-1

Similarly, to build a policy engine, a developer now implements the evaluation interfaces. The policy engine has full access to the plain-text data and policy to implement complex logics but has no I/O capability. It can also request the program's information to verify against the policy's constraints. The evaluation result is returned in a simple Boolean value indicating pass or fail.\looseness=-1

\section{Application Analysis}
\label{sec:application}

In this section, we present an analysis on how to apply our design to the motivating example discussed in \S\ref{sec:design:scenario}, and how \sys can protect the training dataset and output model. Recall that in this use case, multiple hospitals provide confidential diagnostic data to jointly train a cancer classification model in a way that no party has direct access to the model parameters, and they can jointly approve policies to the access of the trained model.

We discuss the policy and consumer program design in details in this section, but we have implemented this scenario as well for feasibility demonstration as well as performance evaluation which we will present in \autoref{sec:performance}.

\mypara{Training Data Preparation.} Each hospital creates a \pad to protect its training diagnostic data. To do this, each hospital needs to use its own producer to generate the \pad and provision the data key received from the producer to a verified key delegator. Here, each hospital is the \textit{custodian} of their own \pads. The key delegator can be offered by a cloud service provider as long as it is verifiable.

\mypara{Jointly Training and Model Policy.} The hospitals agree on a training program as the consumer program using the \term{program-constraints} that takes the input datasets in \pads and output the model protected by \pad as well. The data custodian of the output model is jointly controlled by the hospitals.\looseness=-1

The policy attached to the training data, particularly the \term{input-constraints} can differ based on the considerations of each hospitals. For example, a hospital may require that their contributed training data shares no more than 20\% of the total data; Another hospital may, in contrast, require that its share to be more than 50\% so that it can have a higher quota of the resulting model. Only when all the policies are satisfied can the computation move forward.

The output \pad, which is the cancer classification model, will be attached with a policy generated by the consumer that is agreed by all the hospitals involved. It can be further limited by the \term{output-constraints} of each training data. For example, its policy can specify that the policy may be updated if agreed by entities who jointly control at least $T$ of total share. 

\mypara{Using the Model.} To query or fine tune the model, an organization is limited on the query quota based on a fair usage policy and the policy engine will keep a record of it. It must also use one of the agreed programs specified by the program constraint as the consumer program. The sandboxing mechanism ensures that the model can be used to answer queries without leaking. If the hospital wants to fine-tune the model to fit its own customized needs, it must use an approved fine-tuning program as the consumer program. An output ownership restriction policy can be specified to ensure that the new model is restricted to this hospital, preventing the model being transferred to another entity without the agreement of all hospitals.\looseness=-1

\mypara{Example of Training Data Policy.}
To satisfy the above requirements, an example of the policy attached to the training data can include the following:
\begin{packeditemize}
    \item \textbf{Program Constraint.} The training data can only be used with programs approved by the hospital. The programs can be specified by, for example, hashes and encoded as a raw byte sequence.
    \item \textbf{Input Constraint.} The hospital can limit its contribution to not exceed certain percentage (e.g., 20\%), as a regular integer.
    \item \textbf{Output Constraint.} The program's output, which is the model, must be under the control of the designated entity agreed by all the hospitals, which is a custodian represented by a UUID.
\end{packeditemize}

We present a detailed example of the policy enforcement of this setting in \autoref{apx:case-example}.

\section{Performance Evaluation}
\label{sec:performance}

We evaluated the performance of our complete \sys implementation described in \autoref{subsec:impl-complete}. We first evaluated the above example to show the overheads in the entire workflow of \sys and the scalability of \sys, then we provide a runtime benchmark overhead of the system. Our evaluations were conducted on a Dell PowerEdge R450 with an Intel Xeon Silver 4314 CPU running at 2.4 GHz and 128 GiB of memory among which 2 GiB can be used as the enclave memory.\looseness=-1

To evaluate our system, we first implemented the scenario of hospitals jointly training a cancer classification model discussed in \autoref{sec:application} using the Diagnostic Wisconsin Breast Cancer Database~\cite{cancer-database} to train an SVM model with libsvm~\cite{libsvm}. We use it as one of our subjects and analysed the entire workflow, from input to output, to discuss the overheads of each step in \autoref{subsec:perf-train-workflow}. We used the broken down performance figures with a simulation technique to evaluate the scalability of \sys on a distributed setting. To demonstrate the capability of handling complex applications, we also ported libonnx~\cite{libonnx} and NBench~\cite{NBench:online} to demonstrate the overall performance when a runtime is involved in \autoref{subsec:onnx}.

\begin{figure}
    \centering
    \includegraphics[width=.45\textwidth]{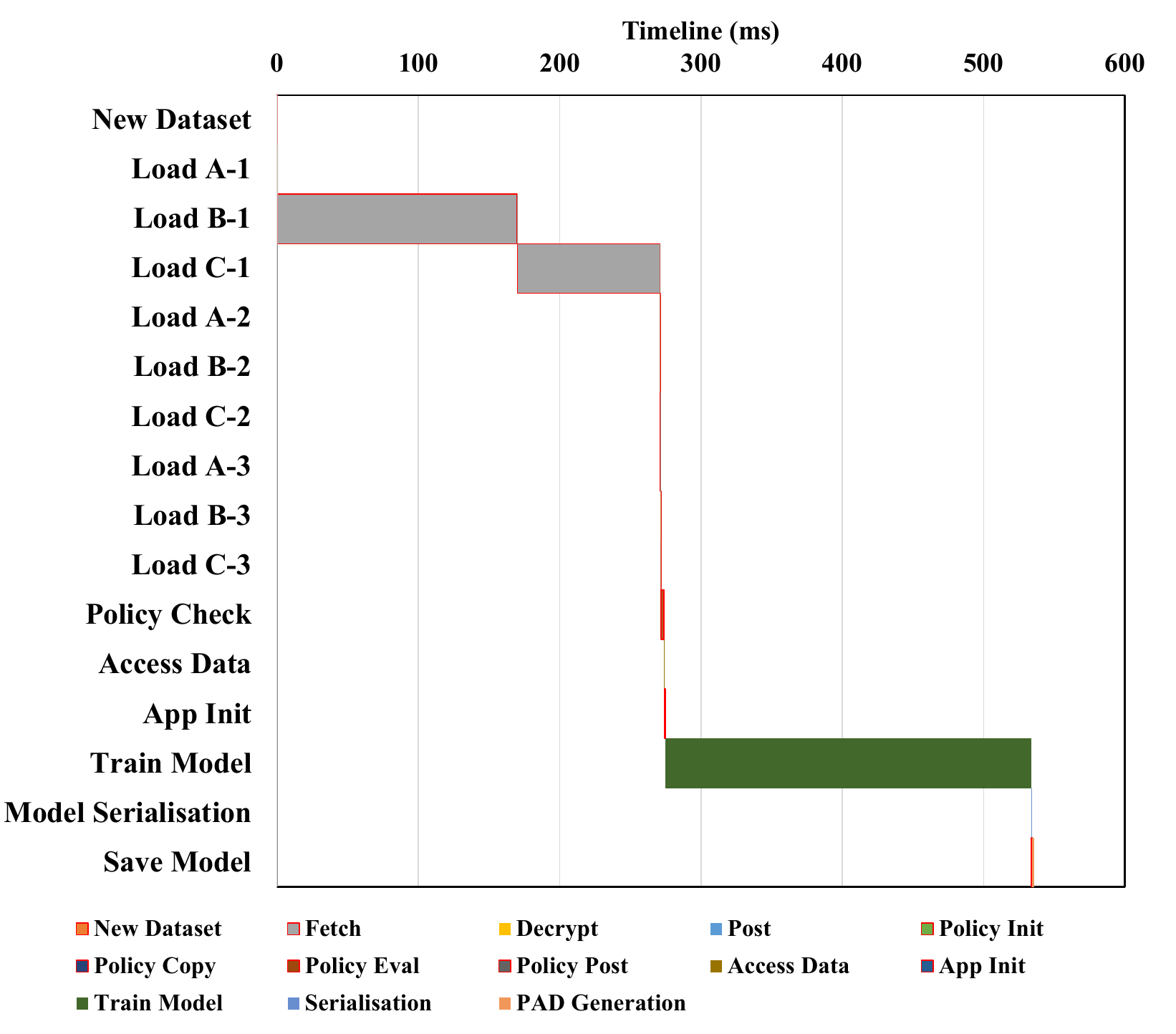}
    \caption{Timeline breakdown of the jointly training example using \sys. Red outline means the overhead is not related to the data's size.}
    \label{fig:svm-timeline}
    \vspace*{-10pt}
\end{figure}

\begin{table}[tb]
\caption{Average overheads breakdown of the framework of \sys measured in ms. Gray overhead title means the overhead is not related to the data's size.}
\small
    \centering
    \resizebox{0.35\textwidth}{!}{
        \begin{tabular}{ll|rr}
\hline\hline

\multicolumn{2}{l|}{}                   & \multicolumn{1}{l|}{w/ Fetching} & w/o Fetching               \\ \hline
\multicolumn{2}{l|}{\textcolor{gray}{New Dataset}}        & \multicolumn{2}{r}{0.007}                                     \\ \hline
\multicolumn{1}{l|}{\multirow{5}{*}{\rotatebox{90}{Load}}} & \textcolor{gray}{Attestation}       & \multicolumn{1}{r|}{135.089} & \multicolumn{1}{r}{-}  \\ \cline{2-4} 
\multicolumn{1}{l|}{} & \textcolor{gray}{Fetch Key}       & \multicolumn{1}{r|}{0.177} & \multicolumn{1}{r}{0.007}  \\ \cline{2-4} 
\multicolumn{1}{l|}{} & Decrypt         & \multicolumn{1}{r|}{0.088}      & \multicolumn{1}{r}{0.098} \\ \cline{2-4} 
\multicolumn{1}{l|}{} & \textcolor{gray}{Policy Matching} & \multicolumn{1}{r|}{0.007}       & \multicolumn{1}{r}{0.008}  \\ \cline{2-4} 
\multicolumn{1}{l|}{} & Total           & \multicolumn{1}{r|}{135.362} & 0.113                     \\ \hline
\multicolumn{1}{l|}{\multirow{5}{*}{\rotatebox{90}{Policy Eval}}} & \textcolor{gray}{Policy Init}        & \multicolumn{2}{r}{1.544}                                 \\ \cline{2-4} 
\multicolumn{1}{l|}{} & Policy Copy      & \multicolumn{2}{r}{0.138}                                    \\ \cline{2-4} 
\multicolumn{1}{l|}{} & Policy Eval        & \multicolumn{2}{r}{0.366}                                   \\ \cline{2-4} 
\multicolumn{1}{l|}{} & \textcolor{gray}{Policy Post}      & \multicolumn{2}{r}{0.025}                                    \\ \cline{2-4}
\multicolumn{1}{l|}{} & Total           & \multicolumn{2}{r}{2.073}                                 \\ \hline
\multicolumn{2}{l|}{Access Data}             & \multicolumn{2}{r}{0.258}                                   \\ \hline
\multicolumn{1}{l|}{\multirow{5}{*}{\rotatebox{90}{Output}}} & \textcolor{gray}{Policy Init}        & \multicolumn{2}{r}{0.119}                                 \\ \cline{2-4} 
\multicolumn{1}{l|}{} & Policy Copy      & \multicolumn{2}{r}{0.074}                                    \\ \cline{2-4} 
\multicolumn{1}{l|}{} & Policy Eval        & \multicolumn{2}{r}{0.330}                                   \\ \cline{2-4} 
\multicolumn{1}{l|}{} & \textcolor{gray}{Policy Post}      & \multicolumn{2}{r}{0.009}                                    \\ \cline{2-4}
\multicolumn{1}{l|}{} & Generate \pad      & \multicolumn{2}{r}{1.405}                                    \\ \cline{2-4}
\multicolumn{1}{l|}{} & Total           & \multicolumn{2}{r}{1.937}                     \\ \hline
\end{tabular}}
    \label{tab:styx-overheads}
\end{table}

\subsection{ML Training Workflow Evaluation}
\label{subsec:perf-train-workflow}

We used the Diagnostic Wisconsin Breast Cancer Database~\cite{cancer-database} to train an SVM model using libsvm~\cite{libsvm} for the workflow overheads evaluation. To provide more insights on the performance of input of \sys, we set 3 hospitals (A, B and C) involving in the training, each provides 3 copies of the database for a total of 9 pieces of data coming from 3 different entities. For each piece of data, the policy discussed in \autoref{sec:application} is attached with proper attributes. The output model is restricted by an \term{output-constraint} in the policy that it must be under the custody of hospital A.

We implemented the policy engine according to the policy language design. The timeline breakdown of the training process is illustrated in \autoref{fig:svm-timeline}. One can easily see that there are 3 events that claimed the most overheads in the workflow: Load B-1, Load C-1 and Train Model. Since the consumer program is owned by hospital A, when first loading dataset of hospital B and C, it will need to contact the key delegator to fetch the data keys which involves communications over sockets and remote attestations. However, once the key is fetched to local, when loading the data from the same entity, \sys no longer needs to communicate with the key delegator and therefore will be much faster. 

The exact value of the overheads are provided in \autoref{tab:styx-overheads}. We can see that most of the overheads are below 2 ms. When fetching the key with the remote attestation and communication involved, we can see that the overheads were around 135 ms, which is comparable to common encrypted transmission methods like HTTPS~\cite{https-latency}.

Note that a red outline in \autoref{fig:svm-timeline} or a gray title in \autoref{tab:styx-overheads} indicates that the overhead is not related to the data's size, meaning that no matter how large the data is, it will always cost the same. We can see that all the overheads that scale with the data's size were lower than 0.4 ms, meaning that \sys can maintain a good scalability and low overhead even with heavy inputs and outputs.

It is worth noting that the overheads of the policy check significantly depend on how complex the logic of the policy engine implements. As a module for the runtime, the performance is also highly related to the runtime's to be discussed in \autoref{subsec:onnx}. For the input and output constraints, the scalability of the policy check is linked to the size of the dataset as each \pad's policy has to be evaluated. For the program constraints, the overhead is static to a specific consumer program and does not increase with the dataset.

\begin{figure}[t]
    \centering
    \includegraphics[width=0.42\textwidth]{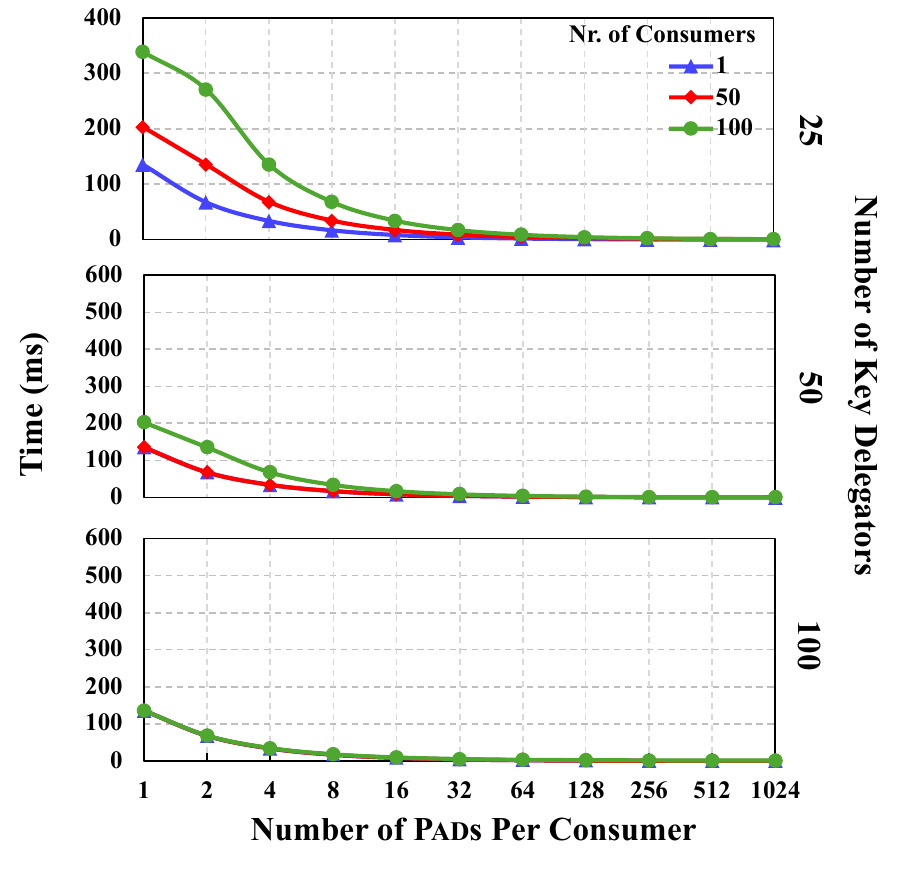}
     \vspace{-0.1in}
    \caption{Per-\pad latency on large scale distributed deployment from simulation.}
    \label{fig:scale-sim}
    \vspace{-0.15in}
\end{figure}

\subsection{Scalability on Distributed Deployment}
\label{subsec:scale}

As a distributed design, \sys is expected to perform stably at large scale. To evaluate the scalability characteristics of \sys, we employed a similar methodology from~\cite{wallet} to simulate the performance of \sys at heavy load with a large amount of nodes. We consider the key delegators behind a simple load balancer that dispatches a key fetching request to whichever key delegator that is free. Once a key delegator has attested a consumer, the upcoming key fetching no longer needs attestation.
We sampled the latency of attestations and key requests with expectation values and standard deviations obtained from our experiments on real machine. In our settings, we simulate a number of consumers trying to decrypt certain amount of \pads all at once to present a heavy load scenario. We simulated with 3 different numbers of key delegators and consumers, with each consumer trying to access different amount of \pads ranging from 1 to 1024.
The results are illustrated in \autoref{fig:scale-sim}. \looseness=-1

From the figure, we can easily observe the trend that the more \pads a consumer needs to decrypt, the lower the latency it eventually needs. This is because the heavy attestation only has to be done once, and the subsequent requests only need the key fetching. 
When the amount of consumers is smaller than the key delegators, the per-\pad latency is capped at a single attestation plus the key fetching latency. This is because every consumer can have a dedicated key delegator serving it. When the amount of consumers is larger than the key delegators, the max latency is in a linear scale to the amount of consumers minus the amount of delegators. 
From the overall results, we can see that even under extreme heavy loads with hundreds of consumers and thousands of \pads to be decrypted, \sys can still achieve a per-\pad latency at max to a couple hundreds of milliseconds.\looseness=-1

\begin{figure*}[th]
    \centering
    \includegraphics[width=.7\textwidth]{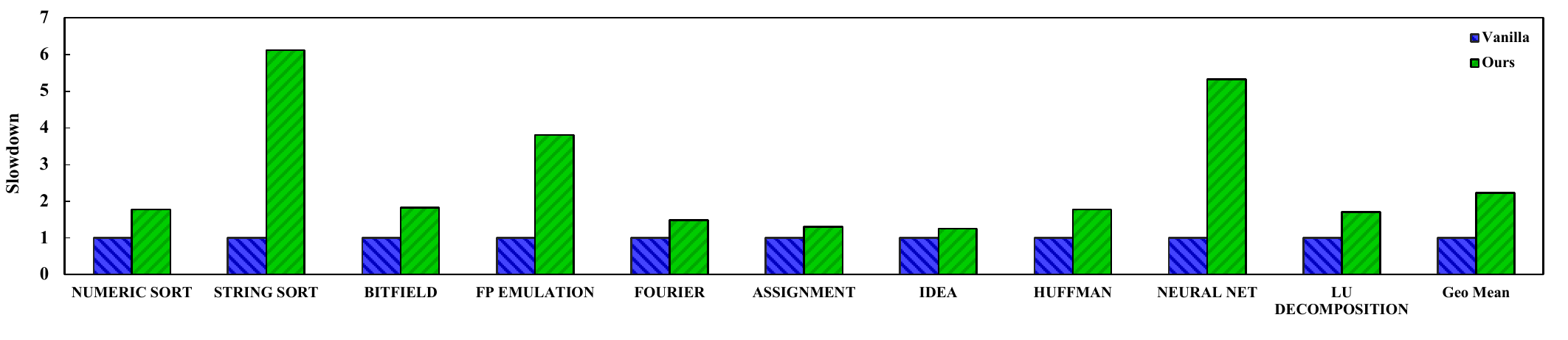}
    \vspace{-0.15in}
    \caption{Normalized NBench slowdown of WAMR over a vanilla result.}
    \label{fig:nbench}
    \vspace{-0.15in}
\end{figure*}

\begin{figure*}[th]
    \centering
    \includegraphics[width=0.90\textwidth]{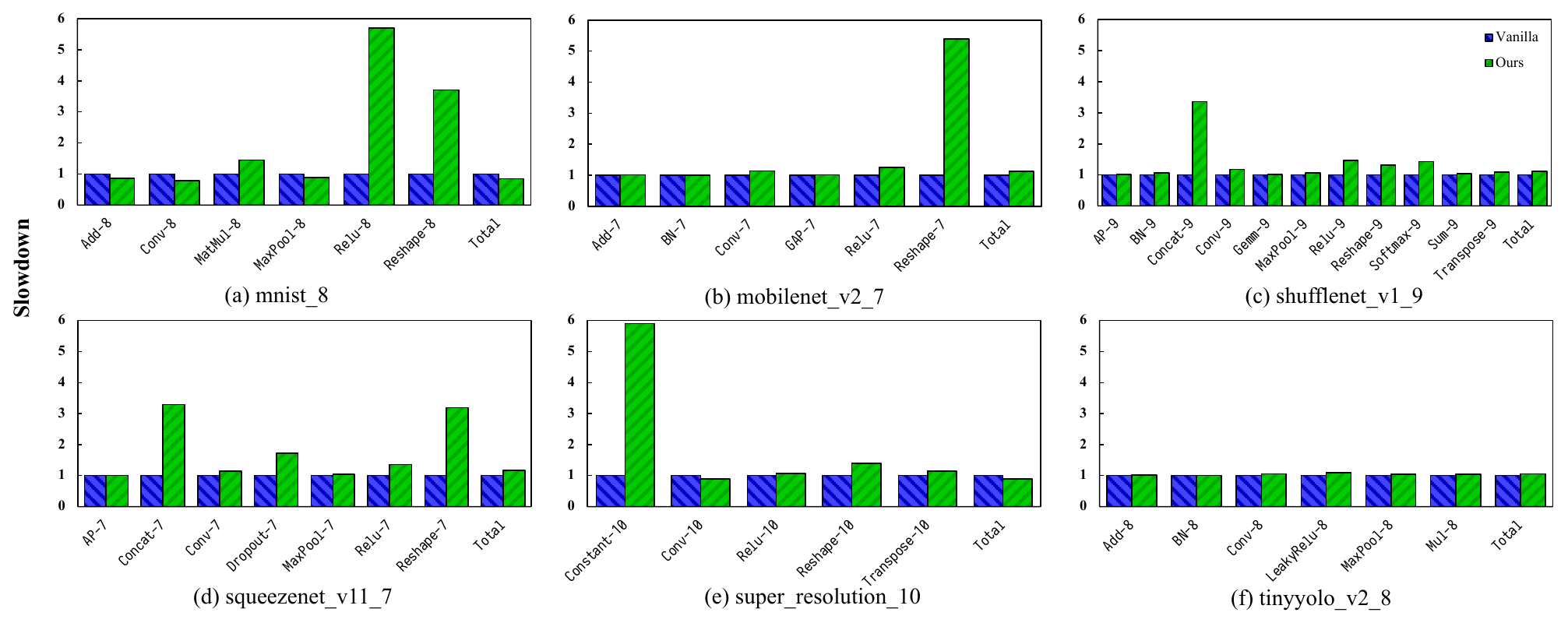}
     \vspace{-0.1in}
    \caption{Normalized {libonnx} benchmark slowdown of the 6 models provided.}
    \label{fig:libonnx-bench}
    \vspace{-0.15in}
\end{figure*}

\subsection{Runtime Performance and Considerations on Complex Applications}
\label{subsec:onnx}

Since \sys involves runtime sandboxing and TEE environments, we present two computation-intensive applications: NBench~\cite{NBench:online} and libonnx~\cite{libonnx} to demonstrate the performance overheads of \sys when using the WAMR runtime. The difference between the two benchmarks is that for NBench, we compiled it into WASM AoT binaries to demonstrate the slow down when heavy computation is done inside the runtime; For libonnx, we ported it as a trusted native binary that the WASM code can call for computation.\looseness=-1

\mypara{Runtime Computation Performance with NBench.} For the vanilla NBench, it was compiled using the GCC with an optimization level of O3. For the WebAssembly version, it was first compiled to WebAssembly using WASI-SDK 20.0~\cite{wasi-sdk:online}, then compiled to AoT binaries, all with an optimization level of O3. The result is normalized to the vanilla's score, that is, how many times slower is presented in \autoref{fig:nbench}. \looseness=-1

As we can see that WAMR imposed slowdown on all benchmarks. While most benchmarks experienced a slowdown around 1.5x, STRING SORT and NEURAL NET suffered the most. However, the geometric mean of the slowdown is around 2.2x which is also close to the benchmark result presented in \cite{wasm-perf:online} (\textasciitilde 2.3x). 

Note that WAMR, as chosen in our system implementation, is just one option for the runtime. Other runtimes can also be used to better suit the characteristics of the workload. Alternatives like Python~\cite{cvspython} and JavaScript (using NodeJS with V8 engine)~\cite{cvsjs} showed different levels of slowdown ranged from 2x as the WASM or up to 20x. A runtime provides the sandboxing required in \sys as well as offers better flexibility and safety for developers. We believe that this is a worthy trade-off over pure C code. \looseness=-1

\mypara{Native Porting with libonnx.} By porting a trusted computation-heavy library natively, we can achieve an almost identical performance to native computation. To demonstrate it as well as presenting the capability of handling complex tasks of \sys, we chose the {libonnx} as our subject. {libonnx} is an execution engine of ONNX~\cite{onnx}, a language for encoding ML models. This means that a jointly-trained model can also be encoded into ONNX format.\looseness=-1

We linked libonnx into the middleware as a trusted native library and tested the performance with {libonnx}'s benchmark suite as the consumer program. The models were packed into \pads with the policies as specified in \autoref{fig:example}. We compared \sys with a vanilla native {libonnx} on  Linux and normalized the benchmark show the slowdown in \autoref{fig:libonnx-bench}. For the total time of the 6 benchmark, the slowdowns are ranging from 0.9x to 1.1x. We can even observe speed-ups in \texttt{mnist\_8} and {\tt super\_resolution\_10}. We suspect this is due to the difference of the memory allocation behavior between the Intel SGX versus a regular Linux. In Intel SGX, heap is pre-allocated so the allocator does not have use syscalls.\looseness=-1

For certain sub-benchmarks showing ``significant'' slowdowns over 3x, the slowdowns are due to the fact that these sub-benchmarks were so fast that their execution time was way smaller than the WASM calling overhead (around 3 us). This made the slowdown looks huge but was not actually impacting the overall performance. Also, the geometric mean of the slowdowns of the 6 benchmarks is 1.03x, making our system on par with a native one when handling tasks using trusted libraries like {libonnx}.\looseness=-1

\mypara{Porting Consideration.} When comparing the two benchmarks, we can see that porting a trusted library natively can greatly help minimizing the overheads. This is suitable for cases where open source libraries are involved. This may also be required when certain external hardware or system features are used that cannot be directly accessed within WASM. However, even with pure WASM code, the overheads were still manageable, allowing more flexibility for developers to implement their propriety code with \sys. \looseness=-1
\section{Discussion} \label{sec:discussion}

\subsection{TCB Scope and Alternative Isolation}

\bheading{TCB Scope.} From our threat model discussed in \autoref{subsec:threat-model} and the architecture introduced in \autoref{sec:design}, we can see that on top of the basic supportive software of the TEE, we added the \sys framework, the runtime and the policy engines into the TCB. For our SGX-based prototype, the basic supportive software is the libraries of SGX SDK.\looseness=-1

\bheading{CVM-Based TEEs.} While our prototype is enclave-based, our design similarly works with confidential VMs (CVMs). Instead of protecting a small piece of software, these CVMs isolates on a VM level granularity. In a similar fashion, we can run the software stack as a process inside the CVM. However, now the supportive software stack will need to include a small OS as a host. A lightweight OS can be sufficient for the purpose such as the one proposed in Gramine-TDX~\cite{gramine-tdx}. When compared to an enclave-based solution, CVM-based TEEs will inevitably increase the size of the TCB.

\bheading{Process Isolation.} TEEs are employed to help defend threats from the outside, particularly against privileged attackers such as the computation infrastructure provider. However, if the infrastructure provider can be trusted, then the only threat will be from the consumer program. In this case, process isolation can be sufficient by running \sys as a process given that the OS it runs upon is trusted.

\subsection{Limitations and Future Works}

While \scheme introduces a novel approach to secure data processing in collaborative environments, our current implementation and design are subject to certain limitations. These areas, however, also offer opportunities for enhancements and further works.\looseness=-1

\bheading{Policy Enforcement in Hardware Accelerators.} Currently, \scheme lacks the capability to enforce policies within hardware accelerators, such as GPUs. This gap is notable as TEE support in GPUs has shown up recently~\cite{nvconfidentialcomputing}. Extending \scheme to integrate with these new TEEs in hardware accelerators represents an important direction for future work, potentially broadening the applicability of \scheme to more tailored computing scenarios, particularly those in AI. \looseness=-1

\bheading{Security-Performance Balance in Runtime Sandboxing.} \sys faces a potential trade-off between security and performance, particularly concerning runtime sandboxing. The practice of compiling bytecode to native machine code ahead-of-time (AoT) for performance improvements may compromise the sandboxing mechanism, as the behaviors of AoT binaries might not be thoroughly validated. However, methods like Software Fault Isolation (SFI) and containerization, as discussed in \cite{reusable-enclaves} and \cite{scone}, provide promising strategies to ensure robust sandboxing even for AoT binaries. The integration of such techniques is left for future investigation.\looseness=-1

\bheading{Middleware Verification.} The consumer middleware of the \scheme framework may benefit from formal verification, which can improve the robustness to the system's overall security~\cite{karim2023blediff,wang2020analyzing,cicala2021pure,deng2018ceive, hussain2018lteinspector}. While formal verification imposes a substantial amount of work, precedents set by projects like \cite{mesapy} has demonstrated the feasibility and value of such efforts. Additionally, the application of SFI, akin to those mentioned in \cite{reusable-enclaves}, could offer interim safeguards for the runtime, even in the absence of formal verification.
\section{Related Works}

\subsection{Sticky Policy Enforcement}

\pad or \emph{sticky policy} is used in \sys to embed policies into data. While recent surveys~\cite{miorandi2020survey,Pearson2011sticky} showed advancement on the sticky policy's design, the enforcement of such policies is the key to the security guarantee. Existing works focus on encryption based techniques. Examples such as Identity-Based Encryption~\cite{boneh2001identity}, Attribute-Based Encryption~\cite{goyal2006attribute,bethencourt2007ciphertext}, and Proxy Re-encryption~\cite{green2007identity} fall into this category. These encryption-based enforcement requires a data consumer to have proper keys that represents the identity or privilege for the access. If all the necessary keys are met under the policy proving the permission, the keys will be able to decrypt the data based on the cryptographic design. \looseness=-1

Compared with \sys, existing works have limitations on the expressiveness of the policy. They also lack the support to impose policies on applications and derived outputs. This means that once the policy check is passed and the data is decrypted, the usage of the data and the outputs will not be in the control of the data owner and therefore cannot satisfy the data-in-use protection we envisioned.\looseness=-1

\subsection{Multi-Stakeholder Computing and ML}

There have been efforts on securing multi-stakeholder computing where stakeholders are mutually distrusted. However, most of the existing works focus on computation configurations and rely heavily on the perfection of the code via verification. For example, \textsc{Pal\ae mon} offers a computation policy enforcement with a combination of human effort certification and TEE reporting~\cite{palaemon}. R. Walther, et al. presented a computation configuration policy enforcement for untrusted cloud~\cite{mspe}. Compared with \sys, the two methods focused on the code of the data consumers and cannot provide flexible per-data policy enforcement throughout the data lifecycle. They also did not address the dynamic collaboration problem. S. Tokuda, et al. proposed a static taint analysis method to enable arbitrary code loading while maintaining certain properties of the data security~\cite{dduc}. This, however, relies on assumptions that the consumer program must be simple enough for the static analysis before loading, limiting what the consumer program can do, not to mention the lack of data lifecycle protection and dynamic collaboration. \looseness=-1

Given our motivating example, multi-stakeholder ML is also related to \sys. Confidential Federated Learning (CFL) is one way to achieve almost full control of the sensitive data locally~\cite{cfl}. However, CFL lacks protection on the output model, and cannot achieve policies such as percentage limit of one entity in the entire training dataset. However, we believe that \sys can be used in combination with existing methods like CFL to achieve such control.

\subsection{Middleware Sandboxing}

The dominant technique for data-in-use protection of \sys is middleware sandboxing. Existing works on middleware sandboxing have explored more on simple access controls with either limited policy and data-in-use protection or limited use cases.

Twine~\cite{twine} offers a WASM sandbox that can host trusted workloads with transparent encrypted I/O. This works in an opposite way comparing to our method where the workload itself is actually untrusted and the I/O of which is strictly limited.

Ryoan~\cite{ryoan} achieved sandboxing via NaCL~\cite{nacl} and controls the I/O via ownership labeling. Data being processed on an entity will have the entity's ownership label added. An entity can remove its label once it has finished the processing and make sure that there is no sensitive data from it is left. Compared with \sys, Ryoan only offers control on ownership level with no possibility to specify complex policies like the inter-\pad rules \sys supports. It also has no policy control on the application and the derived data, meaning that once the data is processed and new data is created, the new data is completely out of the original data owner's control.

\textsc{Laputa}~\cite{laputa} offers private data analysis on a TEE with policy check. It takes a specialized form of code called query plans, and performs policy check with policies defined using regular expressions. Compared with \sys, \textsc{Laputa} cannot execute generic code and can only take the special form of query plan. It also does not take multiple pieces of data like \sys for a collaborative scenario. The policy, while flexible, can only be applicable to the query plan form. There is also no policy control on the derived output, which is expected to contain no sensitive information via the policy check.

When compared with \sys, existing works on middleware sandboxing focus on the input side of a single piece of data and control under what circumstances the data can be read. They, however, do not handle derived policies for derived data, and have limited support on the input and application policies for multiple pieces of data from different stake holders under a collaborative setting.
\section{Conclusions}
We have presented \sys, a novel framework that combines the flexibility of sticky policy and strong enforcement of TEEs to ensure data protection throughout its lifecycle in distributed collaborative settings. \sys achieved data-in-use protection, data lifecycle protection and dynamic collaboration via runtime sandboxing using a middleware. Our application analysis and system implementation demonstrated the feasibility, effectiveness and scalability while imposing reasonable overheads similar to conventional encrypted communication channels even under a distributed multi-node deployment. With this design, \sys can enforce per-data policy from multiple stakeholders throughout the entire data lifecycle for modern collaborative workloads like AI training and other distributed privacy-concerning data processing scenarios.

\section*{Acknowledgments}
We would like to thank the reviewers and shepherd for their feedback and suggestions. The authors from The Ohio State University were partially supported by NSF awards 2207202. The authors from Purdue University were partially supported by NSF awards 2207204. Any opinions, findings, conclusions or recommendations expressed in this material are those of the authors and do not necessarily reflect the views of NSF.

\bibliographystyle{ACM-Reference-Format}
\bibliography{bibs/dcc, bibs/privacy,bibs/Ninghui,bibs/AccessControl,bibs/background}

@inproceedings{karjoth2002platform,
  author    = {G{\"{u}}nter Karjoth and
               Matthias Schunter and
               Michael Waidner},
  title     = {Platform for Enterprise Privacy Practices: Privacy-Enabled Management
               of Customer Data},
  booktitle = {Privacy Enhancing Technologies, Second International Workshop, {PET}
               2002, San Francisco, CA, USA, April 14-15, 2002, Revised Papers},
  series    = {Lecture Notes in Computer Science},
  volume    = {2482},
  pages     = {69--84},
  publisher = {Springer},
  year      = {2002},
}

@inproceedings{mont2003towards,
  author    = {Marco Casassa Mont and
               Siani Pearson and
               Pete Bramhall},
  title     = {Towards Accountable Management of Identity and Privacy: Sticky Policies
               and Enforceable Tracing Services},
  booktitle = {14th International Workshop on Database and Expert Systems Applications
               (DEXA'03), September 1-5, 2003, Prague, Czech Republic},
  pages     = {377--382},
  publisher = {{IEEE} Computer Society},
  year      = {2003},
  url       = {https://doi.org/10.1109/DEXA.2003.1232051},
  doi       = {10.1109/DEXA.2003.1232051},
}

@article{Pearson2011sticky,
  author    = {Siani Pearson and
               Marco Casassa Mont},
  title     = {Sticky Policies: An Approach for Managing Privacy across Multiple
               Parties},
  journal   = {Computer},
  volume    = {44},
  number    = {9},
  pages     = {60--68},
  year      = {2011},
  url       = {https://doi.org/10.1109/MC.2011.225},
  doi       = {10.1109/MC.2011.225},
}

@article{miorandi2020survey,
  author    = {Daniele Miorandi and
               Alessandra Rizzardi and
               Sabrina Sicari and
               Alberto Coen{-}Porisini},
  title     = {Sticky Policies: {A} Survey},
  journal   = {{IEEE} Trans. Knowl. Data Eng.},
  volume    = {32},
  number    = {12},
  pages     = {2481--2499},
  year      = {2020},
  url       = {https://doi.org/10.1109/TKDE.2019.2936353},
  doi       = {10.1109/TKDE.2019.2936353},
}

@article{park2004uconabc,
  author    = {Jaehong Park and
               Ravi S. Sandhu},
  title     = {The UCON\({}_{\mbox{ABC}}\) usage control model},
  journal   = {{ACM} Trans. Inf. Syst. Secur.},
  volume    = {7},
  number    = {1},
  pages     = {128--174},
  year      = {2004},
  url       = {https://doi.org/10.1145/984334.984339},
  doi       = {10.1145/984334.984339},
}

@article{zhang2005usage,
  author    = {Xinwen Zhang and
               Francesco Parisi{-}Presicce and
               Ravi S. Sandhu and
               Jaehong Park},
  title     = {Formal model and policy specification of usage control},
  journal   = {{ACM} Trans. Inf. Syst. Secur.},
  volume    = {8},
  number    = {4},
  pages     = {351--387},
  year      = {2005},
  url       = {https://doi.org/10.1145/1108906.1108908},
  doi       = {10.1145/1108906.1108908},
}

@article{pretschner2006distributed,
  author    = {Alexander Pretschner and
               Manuel Hilty and
               David A. Basin},
  title     = {Distributed usage control},
  journal   = {Commun. {ACM}},
  volume    = {49},
  number    = {9},
  pages     = {39--44},
  year      = {2006},
  url       = {https://doi.org/10.1145/1151030.1151053},
  doi       = {10.1145/1151030.1151053},
}

@inproceedings{goyal2006attribute,
  author    = {Vipul Goyal and
               Omkant Pandey and
               Amit Sahai and
               Brent Waters},
  editor    = {Ari Juels and
               Rebecca N. Wright and
               Sabrina De Capitani di Vimercati},
  title     = {Attribute-based encryption for fine-grained access control of encrypted
               data},
  booktitle = {Proceedings of the 13th {ACM} Conference on Computer and Communications
               Security, {CCS} 2006, Alexandria, VA, USA, October 30 - November 3,
               2006},
  pages     = {89--98},
  publisher = {{ACM}},
  year      = {2006},
  url       = {https://doi.org/10.1145/1180405.1180418},
  doi       = {10.1145/1180405.1180418},
}

@string{CCS = "CCS"}

@string{NDSS = "NDSS"}

@string{AI = "Artif. Intell."}

@string{AI = "Artificial Intelligence"}

@string{ccs = {Proc.~{ACM} Conference on Computer and Communications Security {(CCS)}}}

@string{ndss = {Network and Distributed System Security Symposium ({NDSS})}}

@article{Standaert2010,
  title={Introduction to side-channel attacks},
  author={Standaert, Fran{\c{c}}ois-Xavier},
  journal={Secure integrated circuits and systems},
  pages={27--42},
  year={2010},
  publisher={Springer}
}

@inproceedings{spectre,
    author={Kocher, Paul and Horn, Jann and Fogh, Anders and Genkin, Daniel and Gruss, Daniel and Haas, Werner and Hamburg, Mike and Lipp, Moritz and Mangard, Stefan and Prescher, Thomas and Schwarz, Michael and Yarom, Yuval},
    booktitle={2019 40th IEEE Symposium on Security and Privacy, SP 2019}, 
    title={{Spectre Attacks: Exploiting Speculative Execution}}, 
    year={2019},
    publisher={IEEE}
}

@misc{wasm,
    title={{WebAssembly}},
    author={Web Assembly},
    howpublished={\url{https://webassembly.org}},
    note = {(Accessed on 30/04/2022)}
}

@misc{onnx,
    title={{ONNX | Home}},
    howpublished={\url{https://onnx.ai}},
    author={ONNX},
    note = {(Accessed on 25/01/2024)}
}

@misc{cvspython,
    title={{C VS Python benchmarks}},
    howpublished={\url{https://programming-language-benchmarks.vercel.app/c-vs-python}},
    author={Programming Language and compiler Benchmarks},
    note = {(Accessed on 26/01/2024)},
    year = {2024}
}

@misc{cvsjs,
    title={{C VS Javascript benchmarks}},
    howpublished={\url{https://programming-language-benchmarks.vercel.app/c-vs-javascript}},
    author={Programming Language and compiler Benchmarks},
    note = {(Accessed on 26/01/2024)},
    year = {2024}
}

@misc{nvconfidentialcomputing,
    title={{Confidential Computing | NVIDIA}},
    howpublished={\url{https://www.nvidia.com/en-us/data-center/solutions/confidential-computing/}},
    author={NVDIA},
    note = {(Accessed on 27/1/2024)}
}

@misc{mesapy,
    title={{mesalock-linux/mesapy: A Fast and Safe Python based on PyPy}},
    howpublished={\url{https://github.com/mesalock-linux/mesapy}},
    author={Sun, Mingshen and Rachum, Ram },
    note = {(Accessed on 27/1/2024)}
}

@inproceedings {ryoan,
author = {Tyler Hunt and Zhiting Zhu and Yuanzhong Xu and Simon Peter and Emmett Witchel},
title = {Ryoan: A Distributed Sandbox for Untrusted Computation on Secret Data},
booktitle = {12th USENIX Symposium on Operating Systems Design and Implementation (OSDI 16)},
year = {2016},
isbn = {978-1-931971-33-1},
address = {Savannah, GA},
pages = {533--549},
url = {https://www.usenix.org/conference/osdi16/technical-sessions/presentation/hunt},
publisher = {USENIX Association},
month = nov
}

@inproceedings{nacl,
    author={Yee, Bennet and Sehr, David and Dardyk, Gregory and Chen, J. Bradley and Muth, Robert and Ormandy, Tavis and Okasaka, Shiki and Narula, Neha and Fullagar, Nicholas},
    booktitle={2009 30th IEEE Symposium on Security and Privacy, SP 2009}, 
    title={{Native Client: A Sandbox for Portable, Untrusted x86 Native Code}}, 
    year={2009},
    publisher={IEEE}
}

@inproceedings{laputa,
  author       = {Byeongwook Kim and
                  Jaewon Hur and
                  Adil Ahmad and
                  Byoungyoung Lee},
  title        = {Secure Data Analytics in Apache Spark with Fine-grained Policy Enforcement
                  and Isolated Execution},
  booktitle    = {32nd Annual Network and Distributed System Security Symposium, {NDSS}
                  2025, San Diego, California, USA, February 24-28, 2025},
  publisher    = {The Internet Society},
  year         = {2025},
  url          = {https://www.ndss-symposium.org/ndss-paper/secure-data-analytics-in-apache-spark-with-fine-grained-policy-enforcement-and-isolated-execution/}
}

@inproceedings {reusable-enclaves,
author = {Shixuan Zhao and Pinshen Xu and Guoxing Chen and Mengya Zhang and Yinqian Zhang and Zhiqiang Lin},
title = {Reusable Enclaves for Confidential Serverless Computing},
booktitle = {32nd USENIX Security Symposium (USENIX Security 23)},
year = {2023},
isbn = {978-1-939133-37-3},
address = {Anaheim, CA},
pages = {4015--4032},
url = {https://www.usenix.org/conference/usenixsecurity23/presentation/zhao-shixuan},
publisher = {USENIX Association},
month = aug
}

@misc{google, title={View build provenance ; cloud build documentation; google cloud}, author={Google}, url={https://cloud.google.com/build/docs/securing-builds/view-build-provenance}, journal={Google}, publisher={Google}}

@misc{amazonIAM2011,
  title={AWS Identity and Access Management (IAM)},
  author={Amazon Web Services, Inc.},
  year={2011},
  howpublished={\url{https://aws.amazon.com/iam/}},
}

@book{cramer2015secure,
  title={Secure multiparty computation},
  author={Cramer, Ronald and Damg{\aa}rd, Ivan Bjerre and others},
  year={2015},
  publisher={Cambridge University Press}
}

@inproceedings{bogetoft2009secure,
  title={Secure multiparty computation goes live},
  author={Bogetoft, Peter and Christensen, Dan Lund and Damg{\aa}rd, Ivan and Geisler, Martin and Jakobsen, Thomas and Kr{\o}igaard, Mikkel and Nielsen, Janus Dam and Nielsen, Jesper Buus and Nielsen, Kurt and Pagter, Jakob and others},
  booktitle={Financial Cryptography and Data Security: 13th International Conference, FC 2009, Accra Beach, Barbados, February 23-26, 2009. Revised Selected Papers 13},
  pages={325--343},
  year={2009},
  organization={Springer}
}

@inproceedings{yao1982protocols,
  title={Protocols for Secure Computations},
  author={Yao, Andrew C.},
  booktitle={Proceedings of the 23rd Annual IEEE Symposium on Foundations of Computer Science},
  year={1982},
}

@inproceedings{shinde2017panoply,
  title={Panoply: Low-TCB Linux Applications With SGX Enclaves.},
  author={Shinde, Shweta and Le Tien, Dat and Tople, Shruti and Saxena, Prateek},
  booktitle={NDSS},
  year={2017}
}

@misc{aws-nitro-enclaves,
  title={AWS Nitro Enclaves},
  author={Amazon Web Services, Inc.},
  year={2020},
  howpublished={\url{https://aws.amazon.com/ec2/nitro/nitro-enclaves/}},
}

@misc{aws-kms,
  title={AWS Key Management Service},
  author={Amazon Web Services, Inc.},
  year={2020},
  howpublished={\url{https://aws.amazon.com/kms/}},
}

@misc{Google2020ShieldedVMs,
  title={Google Cloud Shielded VMs},
  author={Google Cloud},
  year={2020},
  howpublished={\url{https://cloud.google.com/security/shielded-cloud/shielded-vm}},
}

@misc{Google2020KMS,
  title={Google Cloud Key Management Service},
  author={Google Cloud},
  year={2020},
  howpublished={\url{https://cloud.google.com/kms/}},
}

@inproceedings{yao1986generate,
  title={How to generate and exchange secrets},
  author={Yao, Andrew Chi-Chih},
  booktitle={Annual Symposium on Foundations of Computer Science},
  year={1986},
}

@inproceedings{goldreich2019play,
  title={How to play any mental game, or a completeness theorem for protocols with honest majority},
  author={Goldreich, Oded and Micali, Silvio and Wigderson, Avi},
  booktitle={Providing Sound Foundations for Cryptography: On the Work of Shafi Goldwasser and Silvio Micali},
  pages={307--328},
  year={2019}
}

@inproceedings{galil1987cryptographic,
  title={Cryptographic computation: Secure fault-tolerant protocols and the public-key model},
  author={Galil, Zvi and Haber, Stuart and Yung, Moti},
  booktitle={Conference on the Theory and Application of Cryptographic Techniques},
  pages={135--155},
  year={1987},
  organization={Springer}
}

@inproceedings{malkhi2004fairplay,
  title={Fairplay-Secure Two-Party Computation System.},
  author={Malkhi, Dahlia and Nisan, Noam and Pinkas, Benny and Sella, Yaron and others},
  booktitle={USENIX security symposium},
  volume={4},
  pages={9},
  year={2004},
  organization={San Diego, CA, USA}
}

@inproceedings{huang2011faster,
  title={Faster secure $\{$Two-Party$\}$ computation using garbled circuits},
  author={Huang, Yan and Evans, David and Katz, Jonathan and Malka, Lior},
  booktitle={20th USENIX Security Symposium (USENIX Security 11)},
  year={2011}
}

@inproceedings{mohassel2017secureml,
  title={Secureml: A system for scalable privacy-preserving machine learning},
  author={Mohassel, Payman and Zhang, Yupeng},
  booktitle={2017 IEEE symposium on security and privacy (SP)},
  pages={19--38},
  year={2017},
  organization={IEEE}
}

@techreport{gentry2009fully,
  author = {Gentry, Craig},
  title = {A Fully Homomorphic Encryption Scheme},
  year = {2009},
  institution = {Stanford University},
  type = {Technical Report},
  number = {STAN-CS-09-XXX},
  url = {https://crypto.stanford.edu/craig/craig-thesis.pdf}
}

@article{brakerski2011efficient,
  author = {Brakerski, Zvika and Vaikuntanathan, Vinod},
  title = {Efficient Fully Homomorphic Encryption from (Standard) {LWE}},
  journal = {SIAM Journal on Computing},
  volume = {43},
  number = {2},
  pages = {831--871},
  year = {2011},
  url = {https://eccc.weizmann.ac.il/report/2011/078/}
}

@article{vaikuntanathan2010computing,
  author = {Vaikuntanathan, Vinod},
  title = {Computing Blindfolded: New Developments in Fully Homomorphic Encryption},
  journal = {Bulletin of the EATCS},
  volume = {100},
  pages = {31--54},
  year = {2010},
  url = {https://eccc.weizmann.ac.il/report/2010/131/}
}

@inproceedings{gentry2011implementing,
  author = {Gentry, Craig and Halevi, Shai},
  title = {Implementing Gentry’s Fully-Homomorphic Encryption Scheme},
  booktitle = {Proceedings of the 31st Annual International Conference on Advances in Cryptology (CRYPTO '11)},
  pages = {129--148},
  year = {2011},
  url = {https://eprint.iacr.org/2011/133.pdf}
}

@inproceedings{dowlin2016cryptonets,
  author = {Dowlin, Nathan and Gilad-Bachrach, Ran and Laine, Kim and Lauter, Kristin and Naehrig, Michael and Wernsing, John},
  title = {CryptoNets: Applying Neural Networks to Encrypted Data with High Throughput and Accuracy},
  booktitle = {International Conference on Machine Learning (ICML)},
  year = {2016},
  url = {http://proceedings.mlr.press/v48/dowlin16.pdf}
}

@article{padget2015policy,
  title={Policy-carrying data: A step towards transparent data sharing},
  author={Padget, Julian and Vasconcelos, Wamberto W},
  journal={Procedia Computer Science},
  volume={52},
  pages={59--66},
  year={2015},
  publisher={Elsevier}
}

@inproceedings{park2002towards,
  title={Towards usage control models: beyond traditional access control},
  author={Park, Jaehong and Sandhu, Ravi},
  booktitle={Proceedings of the seventh ACM symposium on Access control models and technologies},
  pages={57--64},
  year={2002}
}

@inproceedings{boneh2001identity,
  title={Identity-based encryption from the Weil pairing},
  author={Boneh, Dan and Franklin, Matt},
  booktitle={Annual international cryptology conference},
  pages={213--229},
  year={2001},
  organization={Springer}
}

@inproceedings{bethencourt2007ciphertext,
  title={Ciphertext-policy attribute-based encryption},
  author={Bethencourt, John and Sahai, Amit and Waters, Brent},
  booktitle={2007 IEEE symposium on security and privacy (SP'07)},
  pages={321--334},
  year={2007},
  organization={IEEE}
}

@inproceedings{green2007identity,
  title={Identity-based proxy re-encryption},
  author={Green, Matthew and Ateniese, Giuseppe},
  booktitle={Applied Cryptography and Network Security: 5th International Conference, ACNS 2007, Zhuhai, China, June 5-8, 2007. Proceedings 5},
  pages={288--306},
  year={2007},
  organization={Springer}
}

@article{sev2020strengthening,
  title={Strengthening VM isolation with integrity protection and more},
  author={Sev-Snp, AMD},
  journal={White Paper, January},
  volume={53},
  pages={1450--1465},
  year={2020}
}

@inproceedings{xu2004applying,
  title={Applying OM-AM to Analyze Digital Rights Management},
  author={Xu, Shouhuai and Sandhu, Ravi},
  booktitle={7th International Conference on E-Commerce Research},
  volume={53},
  year={2004}
}

@article{asokan2014mobile,
  title={Mobile trusted computing},
  author={Asokan, N and Ekberg, Jan-Erik and Kostiainen, Kari and Rajan, Anand and Rozas, Carlos and Sadeghi, Ahmad-Reza and Schulz, Steffen and Wachsmann, Christian},
  journal={Proceedings of the IEEE},
  volume={102},
  number={8},
  pages={1189--1206},
  year={2014},
  publisher={IEEE}
}

@book{kinney2006trusted,
  title={Trusted platform module basics: using TPM in embedded systems},
  author={Kinney, Steven L},
  year={2006},
  publisher={Elsevier}
}

@article{murdoch2015introduction,
  title={Introduction to Trusted Execution Environments (TEE)—IY5606},
  author={Murdoch, Steven J},
  journal={CiteSeerx: University Park, PA, USA},
  year={2015},
  publisher={Citeseer}
}

@inproceedings{garfinkel2003terra,
  title={Terra: A virtual machine-based platform for trusted computing},
  author={Garfinkel, Tal and Pfaff, Ben and Chow, Jim and Rosenblum, Mendel and Boneh, Dan},
  booktitle={Proceedings of the nineteenth ACM symposium on Operating systems principles},
  pages={193--206},
  year={2003}
}

@inproceedings{vasudevan2012trustworthy,
  title={Trustworthy execution on mobile devices: What security properties can my mobile platform give me?},
  author={Vasudevan, Amit and Owusu, Emmanuel and Zhou, Zongwei and Newsome, James and McCune, Jonathan M},
  booktitle={Trust and Trustworthy Computing: 5th International Conference, TRUST 2012, Vienna, Austria, June 13-15, 2012. Proceedings 5},
  pages={159--178},
  year={2012},
  organization={Springer}
}

@inproceedings{zhang2008security,
  title={Security enforcement model for distributed usage control},
  author={Zhang, Xinwen and Seifert, Jean-Pierre and Sandhu, Ravi},
  booktitle={2008 IEEE International Conference on Sensor Networks, Ubiquitous, and Trustworthy Computing (sutc 2008)},
  pages={10--18},
  year={2008},
  organization={IEEE}
}

@inproceedings{wang2023towards,
  title={Towards Efficient Privacy-Preserving Deep Packet Inspection},
  author={Wang, Weicheng and Lee, Hyunwoo and Huang, Yan and Bertino, Elisa and Li, Ninghui},
  booktitle={European Symposium on Research in Computer Security},
  pages={166--192},
  year={2023},
  organization={Springer}
}

@article{tee-experimental,
title = {An experimental evaluation of TEE technology: Benchmarking transparent approaches based on SGX, SEV, and TDX},
journal = {Computers \& Security},
volume = {154},
pages = {104457},
year = {2025},
issn = {0167-4048},
doi = {https://doi.org/10.1016/j.cose.2025.104457},
url = {https://www.sciencedirect.com/science/article/pii/S0167404825001464},
author = {Luigi Coppolino and Salvatore D’Antonio and Giovanni Mazzeo and Luigi Romano}
}

@inproceedings{gramine-tdx,
author = {Kuvaiskii, Dmitrii and Stavrakakis, Dimitrios and Qin, Kailun and Xing, Cedric and Bhatotia, Pramod and Vij, Mona},
title = {Gramine-TDX: A Lightweight OS Kernel for Confidential VMs},
year = {2024},
isbn = {9798400706363},
publisher = {Association for Computing Machinery},
address = {New York, NY, USA},
url = {https://doi.org/10.1145/3658644.3690323},
doi = {10.1145/3658644.3690323},
booktitle = {Proceedings of the 2024 on ACM SIGSAC Conference on Computer and Communications Security},
pages = {4598–4612},
numpages = {15},
location = {Salt Lake City, UT, USA},
series = {CCS '24}
}

@inproceedings{tnic, author = {Giantsidi, Dimitra and Pritzi, Julian and Gust, Felix and Katsarakis, Antonios and Koshiba, Atsushi and Bhatotia, Pramod}, title = {TNIC: A Trusted NIC Architecture: A hardware-network substrate for building high-performance trustworthy distributed systems}, year = {2025}, isbn = {9798400710797}, publisher = {Association for Computing Machinery}, address = {New York, NY, USA}, url = {https://doi.org/10.1145/3676641.3716277}, doi = {10.1145/3676641.3716277}, booktitle = {Proceedings of the 30th ACM International Conference on Architectural Support for Programming Languages and Operating Systems, Volume 2}, pages = {1282–1301}, numpages = {20}, location = {Rotterdam, Netherlands}, series = {ASPLOS '25} }

@INPROCEEDINGS{raptee,
  author={Pigaglio, Matthieu and Bruneau-Queyreix, Joachim and Bromberg, Yérom-David and Frey, Davide and Rivière, Etienne and Réveillère, Laurent},
  booktitle={2022 IEEE 42nd International Conference on Distributed Computing Systems (ICDCS)}, 
  title={{RAPTEE}: {Leveraging} trusted execution environments for {Byzantine}-tolerant peer sampling services}, 
  year={2022},
  volume={},
  number={},
  pages={603-613},
  doi={10.1109/ICDCS54860.2022.00064}}

@inproceedings {scone,
author = {Sergei Arnautov and Bohdan Trach and Franz Gregor and Thomas Knauth and Andre Martin and Christian Priebe and Joshua Lind and Divya Muthukumaran and Dan O{\textquoteright}Keeffe and Mark L. Stillwell and David Goltzsche and Dave Eyers and R{\"u}diger Kapitza and Peter Pietzuch and Christof Fetzer},
title = {{SCONE}: Secure Linux Containers with Intel {SGX}},
booktitle = {12th USENIX Symposium on Operating Systems Design and Implementation (OSDI 16)},
year = {2016},
isbn = {978-1-931971-33-1},
address = {Savannah, GA},
pages = {689--703},
url = {https://www.usenix.org/conference/osdi16/technical-sessions/presentation/arnautov},
publisher = {USENIX Association},
month = nov
}

@misc{wallet,
      title={Confidential Serverless Computing}, 
      author={Patrick Sabanic and Masanori Misono and Teofil Bodea and Julian Pritzi and Michael Hackl and Dimitrios Stavrakakis and Pramod Bhatotia},
      year={2025},
      eprint={2504.21518},
      archivePrefix={arXiv},
      primaryClass={cs.CR},
      url={https://arxiv.org/abs/2504.21518}, 
}

@inproceedings{mspe,
author = {Walther, Robert and Weinhold, Carsten and Amthor, Peter and Roitzsch, Michael},
title = {Multi-Stakeholder Policy Enforcement for Distributed Systems},
year = {2024},
isbn = {9798400713392},
publisher = {Association for Computing Machinery},
address = {New York, NY, USA},
url = {https://doi.org/10.1145/3702637.3702958},
doi = {10.1145/3702637.3702958},
booktitle = {Proceedings of the 10th International Workshop on Container Technologies and Container Clouds},
pages = {7–12},
numpages = {6},
location = {Hong Kong, Hong Kong},
series = {WoC '24}
}

@article{cfl,
author = {Guo, Jinnan and Pietzuch, Peter and Paverd, Andrew and Vaswani, Kapil},
title = {Trustworthy AI Using Confidential Federated Learning},
year = {2024},
issue_date = {September 2024},
publisher = {Association for Computing Machinery},
address = {New York, NY, USA},
volume = {67},
number = {9},
issn = {0001-0782},
url = {https://doi.org/10.1145/3677390},
doi = {10.1145/3677390},
journal = {Commun. ACM},
month = aug,
pages = {48–53},
numpages = {6}
}

@ARTICLE{twine,
  author={Ménétrey, Jämes and Pasin, Marcelo and Felber, Pascal and Schiavoni, Valerio and Mazzeo, Giovanni and Hollum, Arne and Vaydia, Darshan},
  journal={IEEE Transactions on Dependable and Secure Computing}, 
  title={A Comprehensive Trusted Runtime for WebAssembly With Intel SGX}, 
  year={2024},
  volume={21},
  number={4},
  pages={3562-3579},
  doi={10.1109/TDSC.2023.3334516}}

@InProceedings{dduc,
author="Tokuda, Shota
and Kakei, Shohei
and Shiraishi, Yoshiaki
and Saito, Shoichi",
editor="Song, Houbing Herbert
and Di Pietro, Roberto
and Alrabaee, Saed
and Tubishat, Mohammad
and Al-kfairy, Mousa
and Alfandi, Omar",
title="Decentralized Data Usage Control with Confidential Data Processing on Trusted Execution Environment and Distributed Ledger Technology",
booktitle="Network and System Security",
year="2025",
publisher="Springer Nature Singapore",
address="Singapore",
pages="127--144",
isbn="978-981-96-3531-3"
}

@INPROCEEDINGS{palaemon,
  author={Gregor, Franz and Ozga, Wojciech and Vaucher, Sébastien and Pires, Rafael and Le Quoc, Do and Arnautov, Sergei and Martin, André and Schiavoni, Valerio and Felber, Pascal and Fetzer, Christof},
  booktitle={2020 50th Annual IEEE/IFIP International Conference on Dependable Systems and Networks (DSN)}, 
  title={Trust Management as a Service: Enabling Trusted Execution in the Face of Byzantine Stakeholders}, 
  year={2020},
  volume={},
  number={},
  pages={502-514},
  doi={10.1109/DSN48063.2020.00063}}

@misc{NBench:online,
    author = {Uwe F. Mayer},
    title = {{Linux/Unix nbench.}},
    howpublished = {\url{https://www.math.utah.edu/~mayer/linux/bmark.html}},
    note = {(Accessed on 28/11/2023)}
}

@misc{wasi-sdk:online,
    author = {Andrew Brown},
    title = {{WebAssembly/wasi-sdk: WASI-enabled WebAssembly C/C++ toolchain}},
    howpublished = {\url{https://github.com/WebAssembly/wasi-sdk}},
    note = {(Accessed on 28/11/2023)}
}

@misc{wasm-perf:online,
    author = {Frank Denis},
    title = {{Performance of WebAssembly runtimes in 2023 | Frank DENIS random thoughts.}},
    howpublished = {\url{https://00f.net/2023/01/04/webassembly-benchmark-2023/}},
    note = {(Accessed on 28/11/2023)}
}

@misc{cc-techtarget ,
  author = {Alexanda S Gillis},
  title = {confidential computing},
  month =        Dec,
  year =         "2019",
  url =          "https://www.techtarget.com/searchcloudcomputing/definition/confidential-computing",
}

@misc{cancer-database,
      title = {Breast Cancer Wisconsin (Diagnostic)},
    url =          "https://archive.ics.uci.edu/dataset/17/breast+cancer+wisconsin+diagnostic",
    author = {William, Wolberg and Olvi, Mangasarian and Nick, Street and  W., Street},
    note = {(Accessed on 14/04/2024)}
}

@misc{libsvm,
      title = {cjlin1/libsvm: LIBSVM -- A Library for Support Vector Machines},
    url =          "https://github.com/cjlin1/libsvm",
    author={Chang, Chih-Chung  and Lin, Chih-Jen },
    note = {(Accessed on 14/04/2024)}
}

@misc{libonnx,
      title = {xboot/libonnx: A lightweight, portable pure C99 onnx inference engine for embedded devices with hardware acceleration support.},
    url =          "https://github.com/xboot/libonnx",
    author={Jiang, Jianjun},
    note = {(Accessed on 14/04/2024)}
}

@inproceedings{https-latency, 
author = {Naylor, David and Finamore, Alessandro and Leontiadis, Ilias and Grunenberger, Yan and Mellia, Marco and Munaf\`{o}, Maurizio and Papagiannaki, Konstantina and Steenkiste, Peter}, title = {The Cost of the "S" in HTTPS}, year = {2014}, publisher = {Association for Computing Machinery}, address = {New York, NY, USA}, url = {https://doi.org/10.1145/2674005.2674991}, doi = {10.1145/2674005.2674991}, booktitle = {Proceedings of the 10th ACM International on Conference on Emerging Networking Experiments and Technologies}, pages = {133–140}, numpages = {8}, location = {Sydney, Australia}, series = {CoNEXT '14} }

@misc{cc-arm ,
  author = {Arm Blueprint staff},
  title = {What Is Confidential Computing? Here’s A Great Example},
  month =        Aug,
  year =         "2021",
  url =          "https://www.arm.com/blogs/blueprint/confidential-computing",
}

@INPROCEEDINGS{vsgx,
  author={Zhao, Shixuan and Li, Mengyuan and Zhangyz, Yinqian and Lin, Zhiqiang},
  booktitle={2022 IEEE Symposium on Security and Privacy (SP)}, 
  title={{vSGX}: {Virtualizing SGX Enclaves on AMD SEV}}, 
  year={2022},
  volume={},
  number={},
  pages={321-336},
  doi={10.1109/SP46214.2022.9833694}}

@misc{cc-microsoft ,
  author = {Microsoft team},
  title = {Azure confidential computing: Use cases and scenarios},
  month =        Aug,
  year =         "2022",
  url =          "https://learn.microsoft.com/en-us/azure/confidential-computing/use-cases-scenarios",
}

@misc{cc-ibm ,
  author = {Nataraj Nagaratnam},
  title = {What is confidential computing?},
  month =        Oct,
  year =         "2020",
  url =          "https://www.ibm.com/cloud/learn/confidential-computing",
}

@misc{cc-intel ,
  author = {Ron Perez},
  title = {Case study: Confidential Computing},
  month =        Dec,
  year =         "2021",
  url =          "https://www.intel.es/content/dam/www/central-libraries/us/en/documents/confidential-computing-case-studies.pdf",
}

@string{CRYPTO = "Advances in Cryptology - Annual International Cryptology Conference (CRYPTO)"}

@string{CCS = "ACM Conference on Computer and Communications Security (CCS)"}

@string{ICML = "Proceedings of the International Conference on Machine Learning (ICML)"}

@string{CRYPTO = "CRYPTO"}

@string{ICML = "ICML"}

@inproceedings{wang2020analyzing,
  title={Analyzing the attack landscape of Zigbee-enabled IoT systems and reinstating users' privacy},
  author={Wang, Weicheng and Cicala, Fabrizio and Hussain, Syed Rafiul and Bertino, Elisa and Li, Ninghui},
  booktitle={Proceedings of the 13th ACM Conference on Security and Privacy in Wireless and Mobile Networks},
  pages={133--143},
  year={2020}
}

@inproceedings{deng2018ceive,
  title={Ceive: Combating caller id spoofing on 4g mobile phones via callee-only inference and verification},
  author={Deng, Haotian and Wang, Weicheng and Peng, Chunyi},
  booktitle={Proceedings of the 24th Annual International Conference on Mobile Computing and Networking},
  pages={369--384},
  year={2018}
}

@article{cicala2021pure,
  title={Pure: A framework for analyzing proximity-based contact tracing protocols},
  author={Cicala, Fabrizio and Wang, Weicheng and Wang, Tianhao and Li, Ninghui and Bertino, Elisa and Liang, Faming and Yang, Yang},
  journal={ACM Computing Surveys (CSUR)},
  volume={55},
  number={1},
  pages={1--36},
  year={2021},
  publisher={ACM New York, NY}
}

@inproceedings{karim2023blediff,
  title={BLEDiff: Scalable and Property-Agnostic Noncompliance Checking for BLE Implementations},
  author={Karim, Imtiaz and Al Ishtiaq, Abdullah and Hussain, Syed Rafiul and Bertino, Elisa},
  booktitle={2023 IEEE Symposium on Security and Privacy (SP)},
  pages={3209--3227},
  year={2023},
  organization={IEEE}
}

@inproceedings{hussain2018lteinspector,
  title={LTEInspector: A systematic approach for adversarial testing of 4G LTE},
  author={Hussain, Syed and Chowdhury, Omar and Mehnaz, Shagufta and Bertino, Elisa},
  booktitle={Network and Distributed Systems Security (NDSS) Symposium 2018},
  year={2018}
}

\appendix
\section{Detailed Policy Enforcement Example}
\label{apx:case-example}

Here we present a case analysis with concrete numbers for our motivating example to facilitate a better understanding on the workflow. Suppose we have 3 hospitals: A, B, C. The 3 hospitals forms a joint entity D. There are 3 consumer programs approved by the 3 hospitals: training program, query program and fine-tuning program. Note that the following example serves as a simple example showcasing the workflow.

\subsection{Training Data \pads}

Training data \pads may differ in their input constraints but shares the same program constraints and output constraints:
\begin{packeditemize}
    \item \textbf{Input Constraint.} A cap $c$ on how many percent of data the hospital wants to contribute at max.
    \item \textbf{Program Constraint.} Must use the training program, specified by the hash $h$.
    \item \textbf{Output Constraint.} The output model's custodian $T$ must be D.
\end{packeditemize}

\bheading{Hospital A.} Custodian is A. Its training data \pad has 100 entries. Its input constraint limits that A's data must not be over 60\%.

\bheading{Hospital B.} Custodian is B. Its training data \pad has 50 entries. Its input constraint limits that A's data must not be over 30\%.

\bheading{Hospital C.} Custodian is C. Its training data \pad has 50 entries. Its input constraint limits that A's data must not be over 40\%.

\begin{algorithm}[ht]
	\caption{Policy Engine for Training Data}
	\label{alg:training}
	\begin{algorithmic}[1]
		\Procedure{Training-Input-Eval}{$D$: Set of input \pads, $p_c$: Consumer program}
        \State let $S_D := \sum_{d\in D} d.\text{data\_count}$
		\For{\textbf{each} $d \in D$}
            \If {$\textsc{hash}(p_c) \neq d.\term{program-constraints}.h$}
                \State \textbf{return} \textbf{false}
            \EndIf
            \State let $p := d.\text{data\_count} / S_D$
            \If {$p > d.\term{input-constraints}.c$}
                \State \textbf{return} \textbf{false}
            \EndIf
        \EndFor
        \State \textbf{return} \textbf{true}
		\EndProcedure
        \Procedure{Training-Output-Eval}{$o$: Output model \pad, $D$: Set of input \pads}
        \State let $T := o.\term{data-id}.\text{custodian}$
		\For{\textbf{each} $d \in D$}
            \If {$T \neq d.\term{output-constraints}.T$}
                \State \textbf{return} \textbf{false}
            \EndIf
        \EndFor
        \State \textbf{return} \textbf{true}
		\EndProcedure
    \end{algorithmic}
\end{algorithm}

The code of the policy engine for training is presented in \autoref{alg:training}. We now discuss how these functions performs the policy check in details.

\subsection{Training Data Loading}

When the training program tries to read the training data \pads, it forms the dataset and then request a policy check before the access is granted. The policy evaluation on the input side is in the procedure \textsc{Training-Input-Eval}. This function has 3 possible outcomes:

\bheading{Passing.} The correct training program loads all \pads from A, B and C into a dataset and initiates a policy check. For input constraints, the entire dataset has 50\% of A's data and 25\% of B and C's data each, matching all three \pads's input constraints. The program's hash matches the program constraint's designated hash and therefore the policy check passes, allowing the training program to continue training.

\bheading{Failing on Input Constraints.} Consider the correct training program is only loading \pads from A and B. Now for input constraints, the entire dataset has 66.7\% of A's data and 33.3\% of B's data, violating both A and B's input constraints. The policy check will fail on line 9 and the dataset management module in the middleware will deny the training program from accessing the data.

\bheading{Failing on Program Constraints.} Consider a wrong training program is loaded into the middleware and tries to access the training data \pads of A, B and C. Even though all 3 \pads' input constraints are satisfied, the program's hash will not match the one specified in the program constraints of the \pads. The policy check will fail on line 5.

\subsection{Model Output}

After the training, the training program will attempt to output the model with proposed policies. From the input dataset, all 3 \pads' output constraints require that the output custodian to be D. This means that if the training program assigns the custodian as another entity (e.g., hospital A), the output will be denied. The output policy evaluation is in the procedure \textsc{Training-Output-Eval}.

To ensure that the model can be used for rate-limited query and fine tuning, the following policy is attached to the model's \pad:
\begin{packeditemize}
    \item \textbf{Input Constraints.} 
    \begin{packeditemize}
        \item \texttt{AUTH\_USER} Rule: The model is only accessible when the program's owner is A, B or C, forming a set $O$.
        \item \texttt{RATE\_LIMIT} Rule: For query program, the rate must be limited according to the contribution (e.g., A: 10 queries/min, B: 5 queries/min, C: 5 queries/min), forming a set $R$ for each user.
    \end{packeditemize}
    \item \textbf{Program Constraints.} Must be the query program or the fine-tuning program, specified by a set of hashes $H$.
    \item \textbf{Output Constraints.}
    \begin{packeditemize}
        \item \texttt{CUSTODIAN} Rule: For the fine-tuning program, the new model's custodian $T$ must still be D.
        \item \texttt{SAME\_POLICY} Rule: The new model's input constraints must include the same input, program and output constraints as this model, but may add more constraints.
    \end{packeditemize} 
\end{packeditemize}

\begin{algorithm}[ht]
	\caption{Policy Engine for Model Data}
	\label{alg:model}
	\begin{algorithmic}[1]
        \State let $r_l$ be the current query rate
		\Procedure{Model-Input-Eval}{$m$: Model \pad, $p_c$: Consumer program}
        \If {$\textsc{hash}(p_c) \neq m.\term{program-constraints}.h$}
            \State \textbf{return} \textbf{false}
        \EndIf
        \For {$i \in m.\term{input-constraints}$}
                \Switch {$i$.rule\_type}
                    \Case {\texttt{AUTH\_USER}}
                        \If {$p_c$.owner $\notin i.O$}
                            \State \textbf{return false}
                        \EndIf
                    \EndCase
                    \Case {\texttt{RATE\_LIMIT}}
                        \If {$p_c$ is Query Program}
                            \State Update $r_l$
                            \If {$r_l > i.R[p_c.\text{owner}]$}
                                \State \textbf{return false}
                            \EndIf
                        \EndIf
                    \EndCase
                \EndSwitch
        \EndFor
        \State \textbf{return} \textbf{true}
		\EndProcedure
        \Procedure{Model-Output-Eval}{$o$: Output model \pad, $m$: Input model \pad}
        \State let $T := o.\term{data-id}.\text{custodian}$
        \For {$i \in m.\term{output-constraints}$}
                \Switch {$i$.rule\_type}
                    \Case {\texttt{CUSTODIAN}}
                        \If {$p_c$ is Fine-Tuning Program}
                            \If {$T \neq i.T$}
                                \State \textbf{return false}
                            \EndIf
                        \EndIf
                    \EndCase
                    \Case {\texttt{SAME\_POLICY}}
                        \If {$p_c$ is Fine-Tuning Program}
                            \If {$m.\term{policy} \notin o.\term{policy}$}
                                \State \textbf{return false}
                            \EndIf
                        \EndIf
                    \EndCase
                \EndSwitch
        \EndFor
        \State \textbf{return} \textbf{true}
		\EndProcedure
    \end{algorithmic}
\end{algorithm}

For this policy, the policy engine presented in \autoref{alg:model} is used to evaluate.

\subsection{Querying}

The query program takes an input of the model \pad and a query item (can be a \pad or plaintext data). The model's input constraints and program constraints will be verified using \textsc{Model-Input-Eval}. If the program's owner is not A, B or C, or if the program is not a query program (or a fine-tuning program, but here we consider the query program), the access will be denied on line 4 and 10. If the accessing rate is higher than the input constraint, the access will also be denied on line 16. The result will be available to the owner hospital as there is no extra output constraints on the model. Note that the query only happens when the query program's hash matches the approved one, meaning that the output is indeed a query result. 

\subsection{Fine-Tuning}

Suppose hospital A wants to fine-tune the model with private training data. The fine-tuning program takes the model and the training data as input and outputs a new model. Similar to the querying, if the program's owner is not A, B or C, or if the program is not a fine-tuning program, the access will be denied on line 4 and 10. Here the program's owner is A and the fine-tuning program is the approved one, the access will be granted.

\bheading{Passing Output Constraints Check.} After the fine-tuning, hospital A wants to make sure that the result model is only accessible to itself, so it adds another constraint to the output which is that the model is only accessible when the program's owner is A. The new model's custodian remains as D. The new policy for the output new model \pad will be:
\begin{packeditemize}
    \item \textbf{Input Constraints.} \textit{The model is only accessible when the program's owner is A.} The model is only accessible when the program's owner is A, B or C. For query program, the rate must be limited according to the contribution (e.g., A: 10 queries/min, B: 5 queries/min, C: 5 queries/min).
    \item \textbf{Program Constraints.} Must be the query program or the fine-tuning program, specified by the hashes.
    \item \textbf{Output Constraints.} For the fine-tuning program, the new model's custodian must still be D. The new model's input constraints must include the same input, program and output constraints as this model, but may add more constraints.
\end{packeditemize}
Notice the added policy to the input constraints. As this policy contains all the input model's policy and the custodian is still D, it passes the output constraint check via \textsc{Model-Output-Eval} and is allowed for output.

\bheading{Failing Output Constraints Check.} Suppose A wants to violate the output constraints and allow another hospital E to access the data. To do that, A must modify the input constraints as the model's policy would otherwise only allow the access from A, B, or C. If A merely just append a new \texttt{AUTH\_USER} rule with A and E as the users, the new model is allowed for output but the previous \texttt{AUTH\_USER} rule limiting only A, B and C as the users is still in effective and will deny E's access. If A tries to remove the previous \texttt{AUTH\_USER} rule, it will violate the \texttt{SAME\_POLICY} and the output will be denied on line 36. In either way, no entity outside of A, B and C will be allowed to access this new model. 

Similarly, suppose A wants to remove the rate limit, it also has to remove the \texttt{RATE\_LIMIT} rule in the new model's input constraints, which will violate the \texttt{SAME\_POLICY} and would be denied for output.\looseness=-1

\end{document}